\documentclass[fleqn,usenatbib]{mnras}

\usepackage{newtxtext,newtxmath}
\usepackage[T1]{fontenc}

\DeclareRobustCommand{\VAN}[3]{#2}
\let\VANthebibliography\thebibliography
\def\thebibliography{\DeclareRobustCommand{\VAN}[3]{##3}\VANthebibliography}

\usepackage{graphicx}	% Including figure files
\usepackage{amsmath}	% Advanced maths commands
\usepackage{subcaption}

\usepackage{amssymb}	% Extra maths symbols

\usepackage{xcolor}
\usepackage{rotating}
\usepackage{soul}

\title[QPPs driven by Kink-Unstable Coronal Loops]{Quasi-Periodic Pulsations Driven by Structural Oscillations in a Kink-Unstable Flaring Coronal Loop}

\author[J. Stewart et al.]{
J. Stewart,$^{1}$\thanks{E-mail: jamie.stewart@hse.gov.uk}
P. K. Browning,$^{1}$
M. Gordovskyy$^{2}$
\\
$^{1}$Jodrell Bank Centre for Astrophysics, University of Manchester, Manchester M13 9PL, United Kingdom\\
$^{2}$Centre for Astrophysics Research, University of Hertfordshire, Hatfield AL10 9AB, United Kingdom\\
}

\date{Accepted XXX. Received YYY; in original form ZZZ}

\pubyear{2023}

\begin{document}
\label{firstpage}
\pagerange{\pageref{firstpage}--\pageref{lastpage}}
\maketitle

\begin{abstract}

Twisted coronal loops in the solar atmosphere may become kink-unstable when their magnetic field lines are sufficiently twisted. This instability can trigger magnetic reconnection, leading to the emission of electromagnetic radiation, which manifests as a solar flare. Previous research has demonstrated that oscillations in microwave emissions, resembling observed quasi-periodic pulsations (QPPs), can be generated by the reconnecting loop. Our aim is to investigate the relationship between the oscillations of the loop and these microwave pulsations. Using 3D magnetohydrodynamical simulations, we examine two models: a straight loop in a uniform-density atmosphere and a curved loop in a gravitationally stratified atmosphere. Using new methodology, we extract the reconnecting loop-top from both models and identify structural oscillations. We then compare these oscillations with the gyrosynchrotron (GS) radiation emitted from the simulations, which is forward-modelled using a radiative transfer code. We find that oscillations in the GS emissions are driven by sausage and kink-mode oscillations. However, the relationship between the oscillation frequencies of the GS emission and the identified loop oscillation modes is complex. The dominant mode in the former may result from interference between sausage-mode and kink-mode oscillations or entirely different mechanisms. Results such as these increase our understanding of the time-dependent behaviour of solar flares and lay the groundwork for potential diagnostic tools that could be used to determine physical parameters within a flaring loop.

\end{abstract}


%%%%%%%%%%%%%%%%%%%%%%%%%%%%%%%%%%%%%%%%%%%%%%%%%%

%%%%%%%%%%%%%%%%% BODY OF PAPER %%%%%%%%%%%%%%%%%%

\section{Introduction}

It is widely accepted that solar flares are the manifestation of emitted electromagnetic radiation resulting from a release of stored magnetic energy in complex magnetic structures within the solar corona through magnetic reconnection \citep{Fletcher2011,Benz2017}. Reconnection can be triggered when these structures interact with each other, themselves, or with their surrounding magnetic field \citep{PriestSolarMHD, Shibata2011}. This mechanism rearranges the magnetic field and releases energy. Consequently, hot plasma and accelerated particles emit radiation across the electromagnetic spectrum out of the flaring region. Solar flares are also associated with coronal mass ejections and the release of energetic particles from the sun. They play a significant role in the dynamics of the solar wind and space weather \citep{SolarWind,Vidotto2021}. They can vary in duration, ranging from a few minutes to several hours \citep{Fletcher2011, Benz2017}, and their intensity and frequency of occurrence vary with the solar cycle \citep{Hathaway2010}.\\

Quasi-periodic pulsations (QPPs), short-lived oscillations, are frequently detected in flare emissions in various wavelengths. The earliest documentation of these oscillations can be dated back to a review on solar continuum bursts written by \citet{Thompson1962}. However, it was \citet{Parks1969-wy} who initially drew attention to these oscillations by highlighting a sixteen-second modulation in the X-ray intensity-time profile of a 1968 solar flare. Since then, the presence of QPPs in solar flares has been consistently recorded. In a review by \citet{Inglis2016-ko}, a 30\% detection rate of QPPs was reported in 675 GOES M and X-type flares observed between 2011 and 2016. Subsequently, \citet{Dominique2018} reported a 90\% detection rate of QPPs within the EUV and SXR bands among 90 flares detected during solar cycle 24. Further research has shed light on the properties and potential theoretical mechanisms governing QPP emissions. Statistical studies indicate that QPPs exhibit a range of durations, typically lasting between a few seconds and several minutes \citep{QPPReview2021} with some briefer QPP events \citep{Takakura1983-jf} and longer-lasting oscillations (such as one lasting for over 30 minutes) \citep{Zaqarashvili2013-qv}, being observed. QPPs have been observed in stellar flares \citep{Mitra2005,Mathioudakis2003,Mathioudakis2006} and pre-main sequence star flares \citep{Reale2018}.\\

Advances in QPP detection techniques, outlined by \citet{Broomhall2019}, have unveiled a diverse array of temporal behaviours exhibited by QPPs. These behaviours include aperiodic trends, anharmonic shapes, modulated periods and amplitudes, and QPPs superimposed with background noise. The variability observed in QPP periods and temporal behaviours, coupled with the different electromagnetic signatures they exhibit, hints at the existence of multiple QPP driving mechanisms at play within a flaring region. A comprehensive summary of the current understanding of QPPs, including their observation and their theoretical driving mechanisms, can be found in recent reviews by
\citet{McLaughlin2018}, \citet{Nakariakov2019}, \citet{Doorsselaere2020}, and \citet{QPPReview2021}.\\

Developing our understanding of QPPs holds great potential. Primarily, gaining insight into the driving mechanisms behind QPPs will contribute to a more comprehensive understanding of the time-dependent nature of energy release in flares, an area that has historically not been fully understood. Furthermore, some QPP oscillations have been shown to exhibit a strong correlation with the background parameters of the flaring plasma, such as temperature, magnetic field strength, and plasma density. For example, \citet{Karampelas2023} identified a quantitative relationship between the period of the waves propagating away from a region undergoing oscillatory reconnection and the aforementioned plasma parameters. Oscillatory reconnection has previously been studied as a candidate mechanism for QPP generation \citep{McLaughlin2009, McLaughlin_2012, Thurgood2017, Thurgood_2018, McLaughlin2018, Nakariakov2019, Doorsselaere2020, Stewart_2022, Karampelas2022,Karampelas2023}. As such, this indicates that it is possible to develop seismological tools capable of deducing plasma parameters of a flaring region from QPP data. \\

One important structure related to flares and QPPs is the twisted coronal loop, which can be modelled as a magnetic flux rope. These structures, ubiquitous throughout the corona, are common sources of solar flares \citep{Fletcher2011}. Coronal loops may acquire twist, originating from sunspot rotation or sub-photospheric motions before they emerge from the photosphere, resulting in the formation of a twisted coronal loop \citep{GuidetotheSun,Fan_2009,Archontis_2013}. Oscillations manifest in these loops during reconnection, including kink-mode oscillations, characterised by lateral swaying, sausage-mode oscillations, involving radial expansion and contraction of the loop, and other oscillations such as torsional, fluting, or acoustic modes \citep{Nakariakov2005, De_Moortel_2012,Nakariakov2016}. Previous research points to a potential correlation between these oscillations and the occurrence of QPPs in solar flare data \citep{Nakariakov2003,Li_2020, Kaltman_2023}, though recent observational data suggests that this does not apply to all flaring loops \citep{Shi_2023}.\\

Kink-unstable twisted coronal loops have long been considered a driving mechanism of solar flares and CMEs \citep{Hood1979, Torok2005, Srivastava2010, Kumar2012}. In ideal MHD, this occurs when the twist of the magnetic lines within a magnetic flux rope surpasses a critical value, resulting in a breakdown of equilibrium, loop deformation, and magnetic reconnection, ultimately leading to a flare \citep{Hood1981, Torok2003}. The resultant oscillations from this process are considered candidate mechanisms for QPPs \citep{Gordovskyy2014,Pinto2016, McLaughlin2018, Mishra_2023}. The value of the critical twist depends on various factors, including aspect ratio, plasma and magnetic pressure ratios, and the structure of the surrounding magnetic field \citep{Hood1979, Torok2003, Bareford_2013}. Also significant is the loop's curvature, which can introduce new oscillation modes into the mechanism \citep{Cargill1994,Doorsselaere2009} and affects the stability of the loop \citep{Bareford_2015}. \\

Recently \citet{Smith2022} demonstrated in a simulation of a kink-unstable coronal loop, coupled to a radiative transfer model of microwave emissions, that slowly-decaying microwave oscillations were emitted from the reconnection site irrespective of the inclusion or exclusion of energetic electrons in their gyrosynchrotron radiation calculations. These oscillations, resembling QPPs, may result from a standing global magnetohydrodynamic (MHD) mode modulating the radiation emitted by the reconnecting plasma. While the precise mechanism driving these oscillations remains unidentified, "structural oscillations" (i.e. sausage, kink, torsional modes etc.) are potential candidates. It should be noted that \citet{Smith2022} also identify strong higher frequency quasi-periodic pulsations associated with rapid variations in the electron acceleration process, possibly due to the triggering of anomalous resistivity. This is an example of the generation of QPPs by temporal (and spatial) variations in energetic electron acceleration, which may be a key mechanism for QPPs as found by \citet{Fleishman2008} and \citet{Collier2024}. However, our focus here is on characterising the MHD modes which arise in a reconnecting loop, as well as their potential observable signatures. \\ 

Motivated by these recent findings, we aim to explore the relationship between the structural oscillations of kink-unstable coronal loops and the observed oscillations in emitted gyrosynchrotron (GS) radiation. To this end, we conduct MHD simulations of straight and curved twisted coronal loops undergoing the kink instability. 
We identify the oscillations of the loop and the internal plasma parameters resulting from this process and determine their connection with the emitted radiation. Since our main interest here is the MHD oscillations, we calculate only emission from 
thermal plasma which should  be most strongly correlated with these oscillations. Emission from non-thermal electrons has been  considered by \citet{Smith2022}, but this has a more complex signature including high-frequency pulsations likely associated with time variations in the energy release and acceleration processes which are outside the scope of this paper. We note that purely thermal flares are observed, albeit rarely \citep{Gary1994, Fleishman2015}.\\

Section \ref{method} describes the straight and curved loop models, their implementation within 3D resistive MHD simulations, and the methodologies used for identifying structural oscillations and calculating GS radiation. Results are presented in Section \ref{Results} and discussed in Section \ref{Discussion}, focusing on the effect of curvature and the implications of these results for flares and QPPs.

\section{Methodology}\label{method}

We investigate two models of a kink-unstable coronal loop in conditions representative of the solar corona, using MHD simulations described in Section \ref{MHDEquations}. The first model, introduced in Section \ref{LinearModel}, represents a straight coronal loop within a constant-density environment. This simpler model serves as a basis for understanding the more realistic curved loop model, simulated in a gravitationally stratified atmosphere, discussed in Section \ref{CurvedLoop}. Section \ref{GS-Calculations} explains how the GS radiation is calculated, while Section \ref{StructuralOscillation-Calculations} focuses on the edge-detection algorithm used for identifying sausage and kink mode oscillations.

\subsection{Solving the Resistive MHD Equations} \label{MHDEquations}

\begin{table*}
\makebox[\textwidth][c]
\centering
\caption{Normalisation constants used in LARE simulations, including user-defined values ($L_0$, $B_0$, $\rho_0$) and their derived counterparts. Values used in the straight and curved loop mode simulations are listed.}
\label{Tab:normalisation_constants}
%\resizebox{\columnwidth}{!}{%
\begin{tabular}{|c|c|c|c|}
\hline
Normalisation Constant & Definition & Value (Straight Loop) & Value (Curved Loop) \\ \hline
$L_0$ & Loop-top Radius & $12\times 10^{6}$ m& $4\times10^6$ m \\ [1.5ex]
  $B_0$& Loop-top Magnetic field Strength & $0.02$ T & $0.02$ T \\ [1.5ex]
 $\rho_0$ & Background Coronal Density & $10^{-11}$ {kg m}\textsuperscript{-3} &  $3\times10^{-11}$ {kg m}\textsuperscript{-3} \\ [1.5ex]
 $v_0$& $\frac{B_0}{\sqrt{\mu_0\rho_0}}$ &  $5.64\times10^6$ m s\textsuperscript{-1}&  $3.26\times10^6$ m s\textsuperscript{-1} \\ [1.5ex] 
 $P_0$ & $\frac{B^2_0}{\mu_0}$ &  $318$ Pa & $318$ Pa \\ [1.5ex]
 $t_0$ & $\frac{L_0}{v_0}$  & $2.07$ s & $1.23$ s \\ [1.5ex]
 $j_0$ & $\frac{B_0}{\mu_0L_0}$ & $1.33\times10^{-3}$ {A m}\textsuperscript{-2} & $3.98\times10^{-3}$ {A m}\textsuperscript{-2} \\ [1.5ex]
 $\epsilon_0$ & $v^2_0$ & $3.18\times10^{13}$ {J kg}\textsuperscript{-1} & $1.06\times10^{13}$ {J kg}\textsuperscript{-1}  \\ [1.5ex]
 $T_0$ & $\frac{1.2m_p\epsilon_0}{k_B}$ & $4.62\times10^{9}$ K &  $1.54\times10^{9}$ K \\ [1.5ex]
 $\eta_0$ & $\mu_0L_0v_0$ & $8.51\times10^{7}$ $\Omega$ m & $1.64\times10^{7}$ $\Omega$ m \\ [1.5ex] \hline
\end{tabular}%}
\end{table*}

We solve a form of the resistive 3D MHD equations in the Lagrangian regime, incorporating a viscous force term denoted as $\textbf{f}_{visc}$, which is implemented to capture weak shocks within the system \citep{Arber2001}. The equations can be expressed as follows:

\begin{gather}
    \frac{D \rho}{Dt} + \rho\nabla\cdot\textbf{v}  = 0\textrm{,} \\
    \rho \frac{D \textbf{v}}{D t} = \left ( \nabla \times \textbf{B} \right ) \times \textbf{B} - \nabla P + \textbf{f}_{\textrm{visc}}\textrm{,} \\
    \frac{D \textbf{B}}{D t } = \left ( \textbf{B} \cdot \nabla \right )\textbf{v} - \textbf{B}\left( \nabla \cdot \textbf{v}\right ) - \eta\nabla \times \left ( \nabla \times \textbf{B} \right )\textrm{,} \\
    \frac{D \epsilon}{D t} = -\frac{P}{\rho} \left ( \nabla \cdot \textbf{v} \right ) + \frac{\eta}{\rho} j^2\textrm{,} \\
    P = \rho \epsilon \left ( \gamma - 1 \right )\textrm{.}
\end{gather}

Here, the mass density is denoted by $\rho$, plasma velocity by $\textbf{v}$, magnetic field by $\textbf{B}$, pressure by $P$, magnetic resistivity by $\eta$, specific energy density by $\epsilon$, current density by $j$, and the heat capacity ratio, set to 5/3, by $\gamma$. While thermal conduction and radiation could influence the observational predictions of our model, they are not incorporated into this research as they were not initially considered by \citet{Smith2022}, with whom we are comparing. Equations are expressed in dimensionless form but the results are presented in dimensional form. The latter is necessary for calculating gyrosynchrotron emissions. The normalisation constants, which scale the straight and curved loop models, are defined in Table \ref{Tab:normalisation_constants}.\\

The viscous force term $\textbf{f}_{visc}$ incorporated in our simulations was initially developed by \citet{Caramana1998} and later adapted to be used in MHD by \citet{Arber2001}. This term consists of three contributions. The first contribution involves approximating the fluid as a set of finite volume masses distributed across a staggered grid, following the method proposed by \citet{VonNeumann1950}. The term is calculated by considering  the nonlinear energy exchange that arises from inelastic collisions among these particles. A second linear term is then included to mitigate non-physical oscillations that may occur behind shock fronts. This approach was first introduced by \citet{Landshoff1955}. Finally, a third term is included to account for errors arising from dividing a continuous fluid into finite volume masses. This correction is necessary to prevent inaccurate viscous dissipation calculations due to self-similar isentropic compression, as discussed by \citet{Caramana1998}. \citet{Caramana1998} combined the work of \citet{Christensen1990} and \citet{Benson1993} to achieve this, introducing a term that deactivates the artificial viscosity in smooth regions of the flow. Combined, these three terms effectively capture weak shocks and contribute a value comparable to the kinetic energy density difference between a plasma element at a grid point and its nearest neighbours.\\

The resistive MHD equations are solved using LARE3D, a Lagrangian remap code, developed by \citet{Arber2001}. We apply zero gradient boundary conditions (except for the velocity at $z=0$ for the curved loop model (see Section \ref{CurvedLoop}). We utilise a current-driven anomalous resistivity, in which the resistivity ($\eta$) increases when the current density exceeds a critical value ($j_{\textrm{crit}}$).

\begin{equation}
    \eta(j) = \left\{
        \begin{array}{ll}
            10^{-6}, & \text{if } j \leq j_{\rm{crit}}\textrm{;}  \\
             10^{-3}, & \text{if } j \geq j_{\rm{crit}}\textrm{.}
        \end{array}
    \right.
\end{equation}

The chosen value of $j_{\rm{crit}}$ is specific to each model and is defined in their respective sections.\\

\subsection{Straight Loop Model} \label{LinearModel}

The development of the kink instability and subsequent reconnection in straight twisted loops has been studied extensively \citep{Browning2003, Browning2008, Bareford_2013, Pinto2015, Bareford2015, Hood2016, Snow2017, Reid2018}. We construct a force-free straight twisted loop of length $L = 20$ following the model used in \citet{Hood2009}, which has previously found success in the study of MHD avalanches resulting from interacting kink-unstable coronal loops \citep{Tam2015,Hood2016,Reid2018}. The initial magnetic field for this model, in cylindrical coordinates, is a force-free equilibrium:

\begin{equation}
B_{\theta} = \left\{
\begin{array}{ll}
\lambda r \left ( 1 - r^2 \right )^3 , & \text{if } r < 1{;} \\
0, &\text{if } r \geq 1 {.} 
\end{array} 
\right. \\
\end{equation}
\begin{equation}
    B_y = \left\{
    \begin{array}{ll}
        \sqrt{ 1 - \frac{\lambda^2}{7} + \frac{\lambda^2}{7} \left ( 1 - r^2 \right )^7 - \lambda^2 r^2 \left ( 1 - r^2 \right )^6}, & \text{if } r < 1 {;} \\
        \sqrt{ 1 - \frac{\lambda^2}{7}}, &\text{if } r \geq 1 {.}
    \end{array} \right.
\end{equation}

Here, $\theta$ represents the azimuthal angle in the x-z plane, $r$ denotes the radius from the origin in the x-z plane, and $\lambda$ signifies the degree of twist in the flux rope. The flux rope undergoes the kink instability when $\lambda > \lambda_c$, where $\lambda_c$ stands for the critical twist. This parameter is also constrained by the requirement that $B^2_y$ must remain positive, limiting $\lambda$ to be less than 2.438 \citep{Hood2009}. We select a value, $\lambda = 2.3$, just above the threshold for the ideal kink instability for a loop with a radius-to-length ratio of 0.05. For simplicity, we use a constant density atmosphere instead of a stratified atmosphere, with $\rho$ and $\epsilon$ set to $1.0$ and $0.01$, respectively. We select $j_{\textrm{crit}}$ manually, examining the system before reconnection and selecting a value of $j$ just above the equilibrium value. This value was set to $j_{\rm{crit}} = 5.0$.\\

We use a 3-dimensional grid, bounded by [-3:3, -10:10, -3:3] with $321\times641\times321$ grid points. The dimensions of this model are chosen to later match, as closely as possible, the resolution of the curved loop model. The ratio of the magnetic field at the centre of the flux rope to the background magnetic field is 2.0, while the plasma-beta is 0.01 inside the loop and 0.05 outside the loop.

\subsection{Curved Loop Model} \label{CurvedLoop}

More realistically, coronal loops are curved with their ends rooted in the photosphere, which may affect both their energy release and oscillations. Various models of curved and twisted coronal loops exist and have been previously used to study topics such as their interaction with non-uniform magnetic fields \citep{Reale2016}, factors influencing their critical twist \citep{Titov1999,Torok2004,Torok2005b}, and the release of energy in MHD avalanches \citep{Cozzo2023}. We use the model developed by \citet{Gordovskyy2014}, which has previously been used by \citet{Pinto2016}, \citet{Gordovskyy2017} and \citet{Smith2022} to study the observational signatures of thermal and non-thermal particles in kink-unstable coronal loops and by \citet{Bareford_2015} to investigate the influence of field geometry and various thermodynamic effects on the stability of twisted flux tubes.\\

We begin by initialising an untwisted magnetic field by positioning two magnetic monopoles beneath the numerical domain, given by:

\begin{equation}
    \textbf{B}(t = 0) = B_1 \left ( \frac{\textbf{r} - \textbf{m}_1}{|\textbf{r} - \textbf{m}_1|^3} - \frac{\textbf{r} - \textbf{m}_2}{|\textbf{r} - \textbf{m}_2|^3} \right ).
\end{equation}

Here $\bf{B}$ represents the magnetic field of the flux rope, $B_1$ scales the magnitude field strength of the loop, and $\bf{r}$ is the position vector from the origin. The vectors $\bf{m_1}$ and $\bf{m_2}$ indicate the positions of the two monopoles: $\bf{m_1}$ = (0,$a$,$-h$) and $\bf{m_2}$ = (0,$-a$,$-h$). The parameter $a$ corresponds to the position of the footpoints on the photosphere. The depth of the monopoles beneath the domain is represented by $h$. We use the values $B_1 = 50$, $a = 6.4$, $h = 3.2$.\\

The twist is created by applying slow vortical motions to each of the loop's circular footpoints. We utilise  the method developed by \citet{Bareford_2015}, which injects twist similar to \citet{Gordovskyy2014} but at a consistent rate that prevents twist dissipation.
The azimuthal velocity within each circular footpoint region is given by 
\begin{gather}
    v_{\rm{rot}}(r,t) =  \psi(r)\zeta(t),\\
    \psi(r) = r\left[ 1 - \tanh\left(\frac{r - R}{\chi}\right)\right] \label{spatial_velocity},\\ 
    \zeta(t) = \frac{w_{\rm{twist}}}{2}\tanh\left(\frac{t - t_1}{\tau_1}\right)\tanh\left(0.5 - \frac{t - t_2}{\tau_2}\right) \label{temporal_velocity}.
\end{gather}

Equation \ref{spatial_velocity} describes the spatial distribution of the twisting motions, where the $r$ represents the radius measured from the centre of the footpoint, while $R$ is the radius of the footpoint. The rotational velocity gradually increases from the centre of the footpoint towards its edge, declining sharply to zero near the footpoint edge. The position of the peak and the rate of decrease is dependent on the parameter $\chi$. We have chosen $R = 0.5$ and $\chi = 0.05$, resulting in the rotational velocity peaking very close to $r = R$ and exhibiting a steep decline thereafter.\\

The temporal evolution of the azimuthal velocity  is described by Equation \ref{temporal_velocity}. Here, $w_{\textrm{twist}}$ scales the magnitude of the rotational velocity. The parameters $t_1$ and $\tau_1$ determine the onset time and the rate at which the twist increases, respectively. Similarly, $t_2$ and $\tau_2$ determine the end time and the rate at which the twist decreases. The factor of 1/2 is included because the twisting profile is applied to both footpoints, which effectively doubles the rate of twist. We have chosen the following parameter values: $w_{\textrm{twist}} = 0.02$, $t_1 = 120$, $t_2 = 460$, $\tau_1 = 40$, and $\tau_2 = 20$, selected to prevent significant dissipation, yet remain slow enough so that the field prior to instability onset is close to equilibrium. Once the loop becomes unstable, at a critical twist of about $4\pi$, we switch off the rotational velocity. The described method of generating a twisted loop by injecting helicity at the footpoints can also be found in studies such as those performed by \citet{Reale2016}, \citet{Reid2018} and \citet{Cozzo2023}. It generates an approximate twisted force-free equilibrium for a curved loop; however, it should be noted that the initial twisting phase in these simulations are not intended to  accurately represent the formation of a real twisted loop.\\

Using the aforementioned model, we construct an untwisted loop within a $512\times512\times512$ Cartesian grid bounded by [x = -10:10, y = -10:10, z = 10:10]. The initial configuration of the loop has a height of 8.34, a footpoint separation of 12.8, a length of 24, and a cross-section at the loop-top of 0.63 in dimensionless units. The magnetic field strength at the loop-top is evaluated to be 0.28, while the magnitude at the footpoint is determined as 4.82. Consequently, we observe an aspect ratio of 38.4 between the length of the loop and the cross-section at the loop-top, as well as a magnetic field strength ratio of 0.058 between the loop-top and the footpoints.\\

We construct a gravitationally-stratified atmosphere, following \citet{Gordovskyy2014}, with three layers: a chromospheric layer situated at the lower boundary of the domain, a transitional layer, and a coronal layer occupying the majority of the domain. The density profile is given by:

\begin{equation}
    \rho(z) = \rho_1 e^{-\frac{z-zc}{z_1}} + \rho_2 e^{-\frac{z-zc}{z_2}}.
\end{equation}.

Here, $\rho_1$ denotes the density of the chromosphere, $\rho_2$ corresponds to the density of the solar corona, $z_c$ represents the height of the transitional layer, and $z_1$ and $z_2$ the gradient of the density in the transitional layer and chromospheric layer respectively. This is consistent with empirical models, such as those discussed by \citet{Vernazza1981}. We set $\rho_1 = 5.15 \times 10^7$, $\rho_2 = 3.03$, $z_c = 4.675\times 10^{-7}$, $z_1 = 5.5\times 10^{-8}$, and $z_2 = 5.0\times 10^{-6}$, with a factor of $10^4$ difference between the density of the chromospheric and coronal layers. These parameters result in a temperature of $ \sim 10^5$ K at the chromospheric level and $ \sim 10^7$ K  at the coronal level. The plasma-beta starts at 0.1 at the footpoints, decreases to 0.005 in the lower corona, and rises to 0.01 at the loop-top.\\

To determine the critical current, we adopt the criterion employed by \citet{Gordovskyy2014},  assuming plasma instabilities leading to increased resistivity arise when the electron drift velocity surpasses the sound speed, i.e., $v_{\text{drift}} > v_{\text{thermal}}$. Consequently, the critical current is expressed as:

\begin{equation}
    j_{\rm{crit}}(\textbf{r}) = \frac{2e}{m_p}\sqrt{\gamma(\gamma - 1)}\rho(\textbf{r})\sqrt{\epsilon(\textbf{r})}.
\end{equation}

\begin{figure*}
    \centering
    \includegraphics[width = 0.7\linewidth]{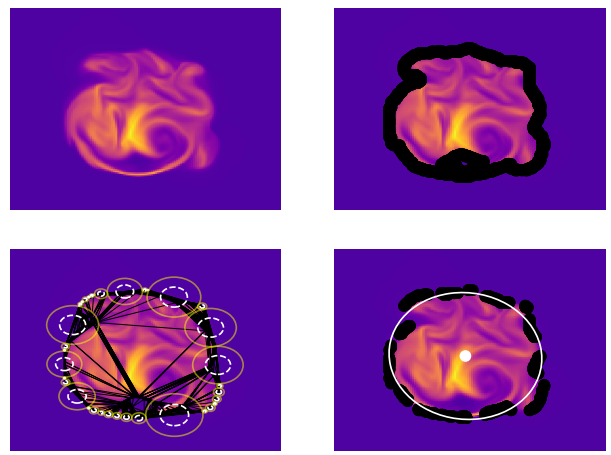}
    \caption{Demonstration of the edge-detection algorithm in action. The top-left image shows an unprocessed input image (a 2D slice of pressure at the loop-top within the midplane of an evolving straight loop), while the top-right image displays the input image with outer edges detected (along with some inner edge artifacts). The bottom-left image illustrates Delaunay triangulation, with black triangles representing the triangles plotted between each point in the edge dataset. The solid yellow circles depict the circumcircles of the edge points associated with the convex hull, and the white dashed circles outline the alpha-shape of these circumcircles. The bottom-right image exclusively features the outer edge points and a fitted ellipse.}
    \label{fig:edge-detection-method}
\end{figure*}

Here, $\textbf{r}$ denotes the position vector, $e$ represents the charge of an electron, and $m_p$ is the mass of a proton. However, as discussed by \citet{Gordovskyy2014}, it is crucial to consider that the current density in global MHD models is limited by the grid resolution. To address this limitation, we multiply the above equation by a factor of $\delta L / R_L$, where $\delta L$ denotes the grid resolution and $R_L$ is the Larmor radius of a proton. In dimensionless units, the critical current is:

\begin{equation}
j_{\rm{crit}} = \frac{2eN}{m_p} \sqrt{\gamma(\gamma-1)\mu_0\rho_0},
\end{equation}

where $N$ is the number of grid points, typically in the direction with the lowest resolution.

\subsection{GS Radiative Transfer Code}\label{GS-Calculations}

Mildly-relativistic electrons within a coronal loop gyrate in magnetic fields leading to GS radiation emission. In a solar flare, this is typically in the microwave frequency range. However, accurately calculating GS radiation is computationally expensive. To overcome this, we use a fast GS radiative transfer code developed by \citet{Fleishman2010}, \citet{Nita2015}, and \citet{Kuznetsov_2021}.\\

The GS code enables the user to take the number density (cm$^{-3}$), temperature (K), and magnetic field  (T) along a line-of-sight and calculate the GS radiation intensity (in solar flux units) emitted along that line-of-sight for a range of selected frequencies. It reduces the computational time required for calculating GS radiation by several orders of magnitude, yielding results within 1-10\% of their exact solutions. The algorithm has previously been implemented in the study of solar flares, \citep{Kontar2017, Gordovskyy2017, Chen2020}, for investigating QPPs \citep{Mossessian2012,Altyntsev2016, Kupriyanova_2022, Smith2022, Kaltman_2023,Shi2023}, and has found applications outside of solar physics \citep{Waterfall2018,Climent2022}.\\

We do not consider non-thermal electrons which has been done by \citet{Smith2022}. Instead, we use a thermal energy distribution and isotropic pitch-angle distribution and compute the radiation emitted along multiple line-of-sights for both models. This focuses on oscillations associated with the MHD behaviour of the loop and allows us to determine how the emitted radiation evolves and identify periods of any fluctuating components that may be correlated with structural or parameter oscillations within the reconnecting loop, increasing our understanding of what mechanisms drive QPPs.

\subsection{Structural Oscillation Analysis}\label{StructuralOscillation-Calculations}

We detect structural oscillations (oscillations of the loop structure, such as sausage modes, kink-modes, etc.) by introducing a new method to isolate structures within a background plasma. Specifically, we isolate the loop-top of both models and study the MHD oscillations occurring therein. To achieve this, we construct a multi-stage algorithm that utilises Canny edge detection \citep{Canny1986}, Delaunay triangulation, and the construction of alpha-shapes to determine the boundaries of structures in a 2D colour map. Sausage-mode and kink-mode oscillations can then be detected by fitting ellipses to the edges of the structure and tracking their evolution over time. A visual demonstration of this algorithm is provided in Figure \ref{fig:edge-detection-method}. This algorithm is not limited to identifying structural oscillations and may lend itself to additional potential applications.\\

The first step in this algorithm is to calculate the edges within a selected 2D slice. We focus on oscillations at the loop-top so we take a 2D slice of the loop's midplane, corresponding to a parameter with a well-defined boundary between the inside and outside of the loop, in this case, pressure. We then remove all values within the image below a threshold value. This results in a crude extraction of the loop-top from the background plasma, which we will further refine. From there we implement the Canny edge detection algorithm to calculate the edges within the 2D image. The Canny edge detection algorithm \citep{Canny1986} has been applied to a variety of non-astrophysical scenarios \citep{Agaian2009,Hou2009}, and has seen continued development in the field of computer vision \citep{Rong2014}. In this paper, we use a traditional method outlined in \citet{CannyBook}.\\

The Canny edge-detection algorithm works as follows: first, we apply a grayscale transformation to the image and then blur it with a Gaussian kernel. This step helps minimise noise within the image. Subsequently, the edges of the image are identified by calculating the gradient of the image. This generates regions with sharp gradients ("strong edges") over a limited number of pixels and larger regions with more gradually changing gradients ("weak edges"). Our final image should have clearly defined edges, so the next step is to convert weak edges into strong edges using non-maximum suppression. This technique involves evaluating for each pixel whether its gradient serves as the local maximum within a neighbourhood of pixels sharing the same gradient direction. If this criterion is met, the local maximum is retained along with any immediately adjacent weak edges, thereby forming a strong edge and enhancing the image's clarity.\\

The outcome is an image containing the extracted edges of the loop-top, with some inner edges left over that we wish to remove to accurately calculate the structural oscillations of the loop. We build upon the Canny edge-detection algorithm by incorporating Delaunay triangulation and implementing alpha-shapes to achieve this. We start by constructing a concave hull around the boundary edges of the isolated structure. To achieve this, we use Delaunay triangulation to generate a set of non-overlapping triangles from the edge dataset. Subsequently, a convex hull is computed around these triangles, providing a preliminary estimate of the boundary between the loop-top and any remaining plasma that had not been removed earlier in the algorithm.\\

Then, we enhance the accuracy of the hull using alpha-shapes. For each point along the convex hull, we compute a circumcircle around the associated triangle's vertices. The circle's radius is then scaled by a parameter, $\alpha$, resulting in the creation of an alpha-shape. Points within the edge dataset that fall within this alpha shape are designated as "boundary edges," and are separated from the interior edges. To create an accurate hull around the loop-top, the $\alpha$ parameter must be carefully chosen. A larger $\alpha$ value produces a less-detailed convex boundary, whereas a smaller value risks missing potential points along the boundary. In our analysis we opted for $\alpha = 0.5$ for the straight and curved loop.\\

The structure is now isolated from the background plasma and can be used for other purposes if desired. To identify structural oscillations, we fit an ellipse to the isolated loop-top for each timestep. By tracking the evolution of the fitted ellipse, we can isolate kink-mode oscillations (through the motion of the elliptical centre) and sausage-mode oscillations (through changes in the area of the ellipse) in the reconnecting loop.

\begin{figure*}
  \centering
  \begin{sideways}
    \begin{minipage}{\textheight} % Use minipage to limit the width
      \centering
      \begin{subfigure}[b]{0.45\textwidth}
        \centering
        \includegraphics[width=\linewidth]{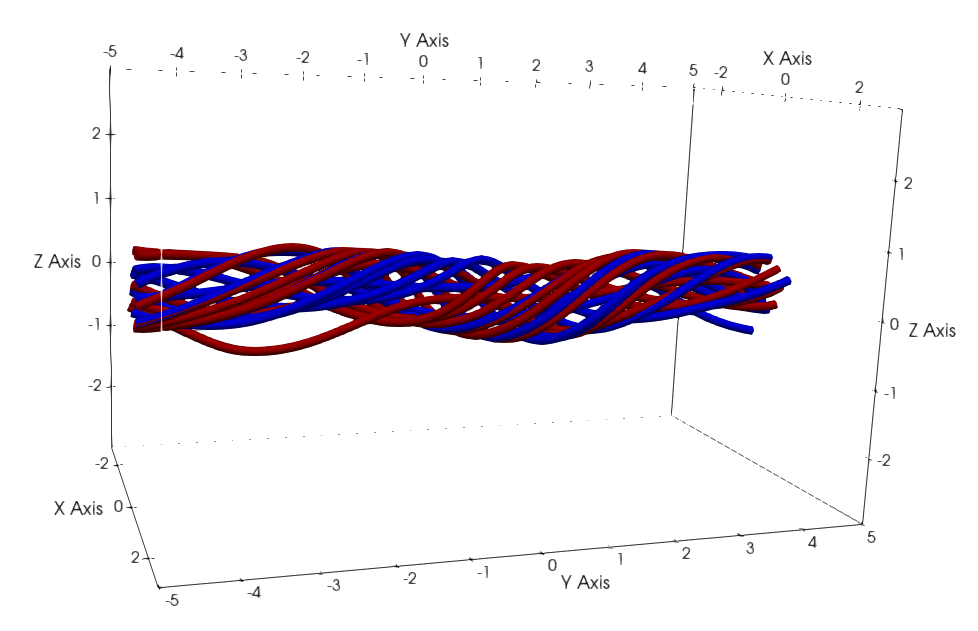}
        \caption{$t = 149.0$ s}
        \label{fig:subfig-a}
      \end{subfigure}
      \hspace{1cm} % Adjust horizontal spacing between subfigures
      \begin{subfigure}[b]{0.45\textwidth}
        \centering
        \includegraphics[width=\linewidth]{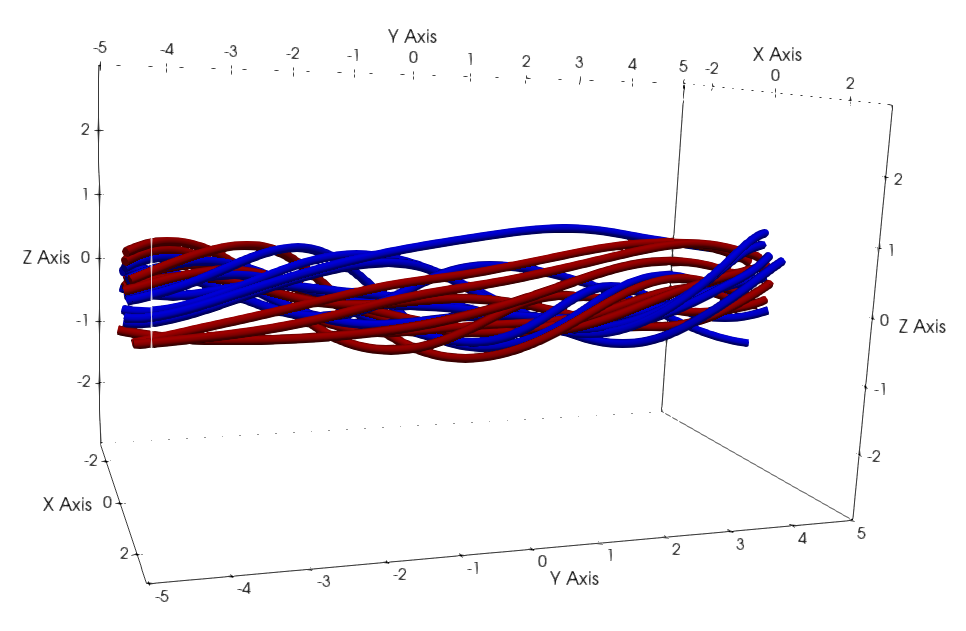}
        \caption{$t = 198.5$ s}
        \label{fig:subfig-b}
      \end{subfigure}
      
      \vspace{1cm} % Adjust vertical spacing between rows
      
      \begin{subfigure}[b]{0.45\textwidth}
        \centering
        \includegraphics[width=\linewidth]{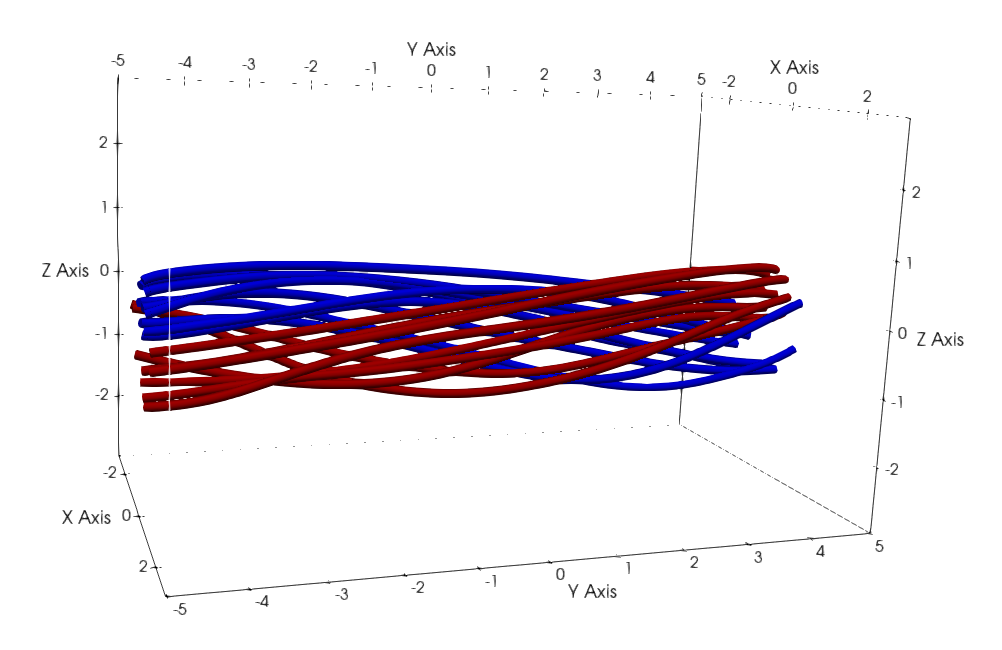}
        \caption{$t = 285.3$ s}
        \label{fig:subfig-c}
      \end{subfigure}
      \hspace{1cm} % Adjust horizontal spacing between subfigures
      \begin{subfigure}[b]{0.45\textwidth}
        \centering
        \includegraphics[width=\linewidth]{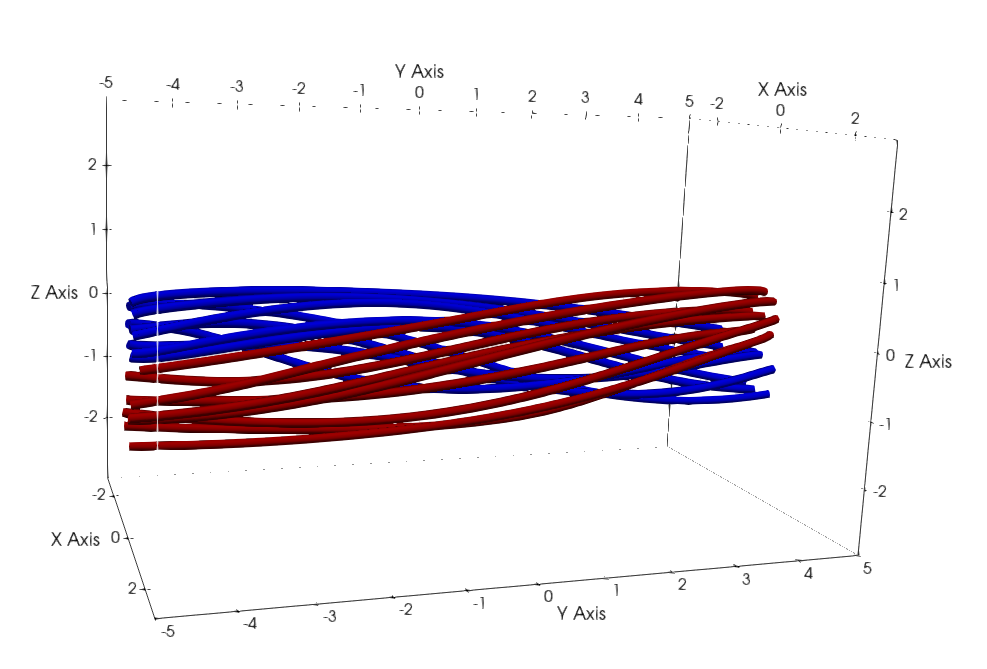}
        \caption{$t = 496.2$ s}
        \label{fig:subfig-d}
      \end{subfigure}
      
      \caption{The evolution of the straight loop's interior magnetic field lines. The blue field lines originate from the foot-point at $y = -10.0$, while the red field lines originate from the foot-point at $y = 10.0$.}
      \label{fig:straight-side-view}
    \end{minipage}
  \end{sideways}
\end{figure*}

\begin{figure*}
    \centering
    \includegraphics[width = \textwidth]{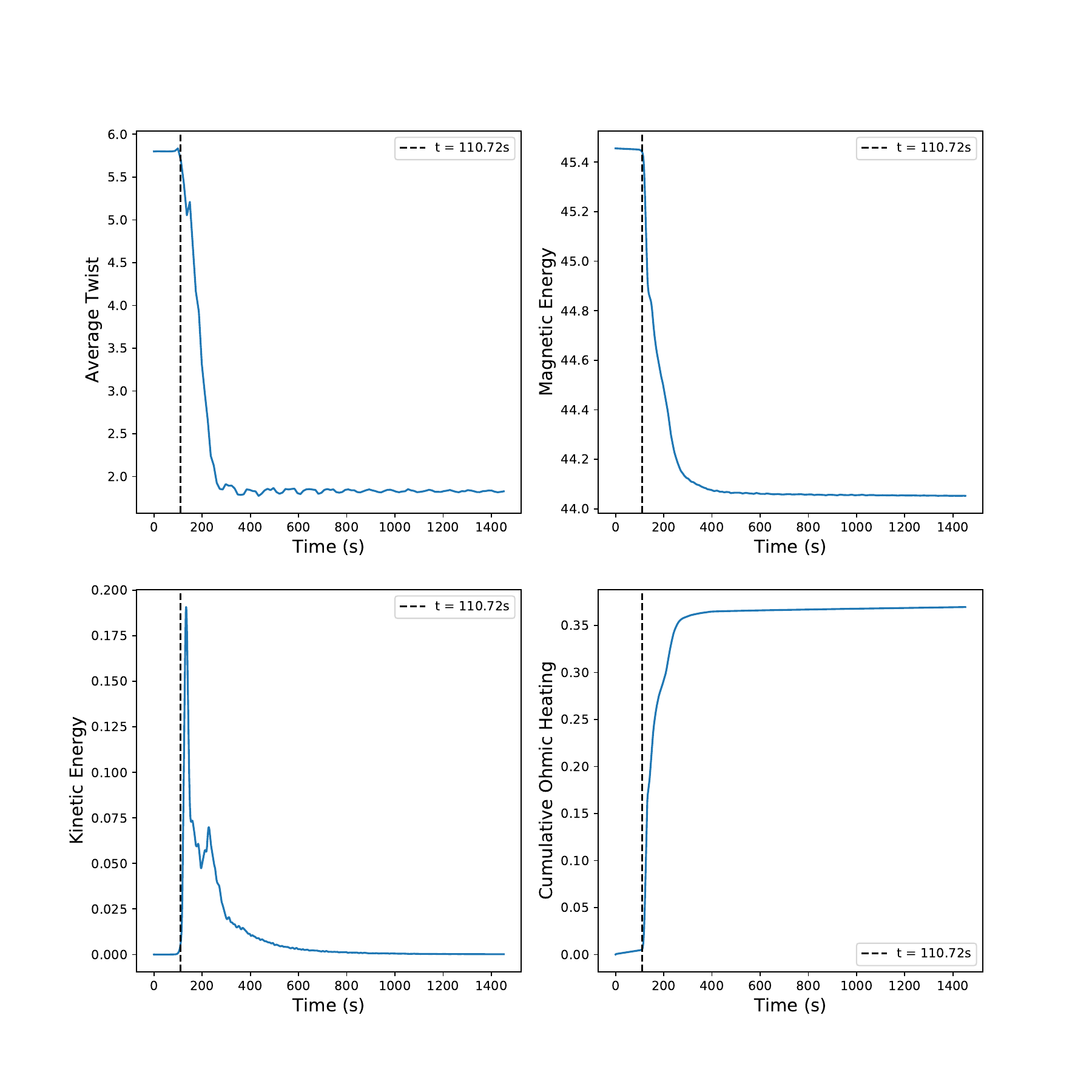}
    \caption{The temporal evolution of the average twist (top-left), total magnetic energy (top-right), kinetic energy (bottom-left) if the system, and cumulative Ohmic heating (bottom-right) for the straight loop. A black dashed line has been marked on each graph at the time at which the total magnetic energy starts to drop, differentiating between different phases of the loop's evolution.}
    \label{fig:straight-energetics}
\end{figure*}

\begin{figure*}
  \centering
  \begin{sideways}
    \begin{minipage}{\textheight} % Use minipage to limit the width
      \centering
          \includegraphics[width=0.9\textwidth, height = 0.9\textheight, keepaspectratio]{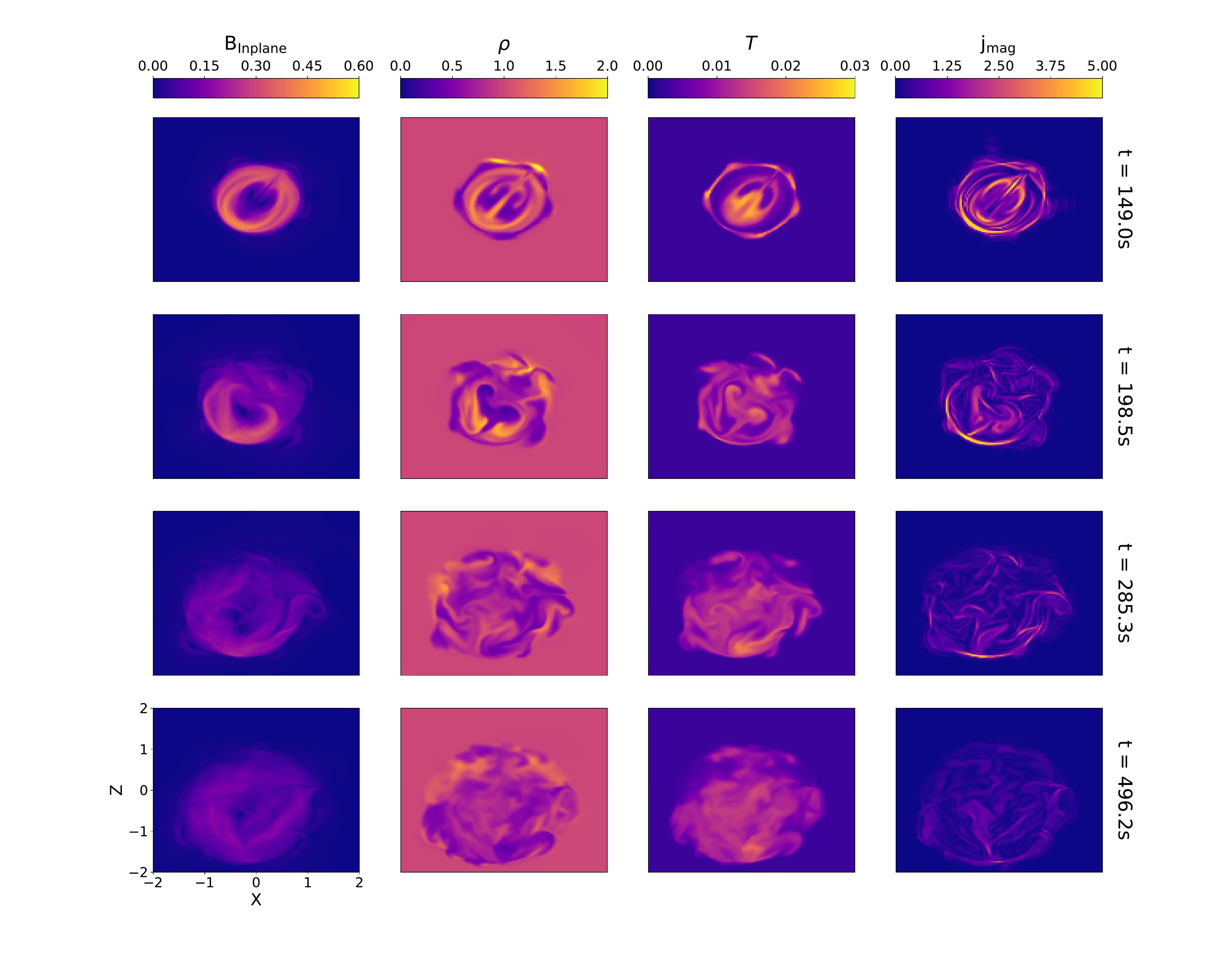}
    \caption{Colour maps illustrating the evolution of the in-plane magnetic field, density, and temperature of the straight loop in the X-Z plane at the midplane of the loop ($y=0$) over time.}
    \label{fig:straight-loop-evolution}
     
    \end{minipage}
  \end{sideways}
\end{figure*}

\begin{figure*}
  \centering
  \begin{sideways}
    \begin{minipage}{\textheight} % Use minipage to limit the width
      \centering
          \includegraphics[width=0.8\textwidth, height = 0.8\textheight, keepaspectratio]{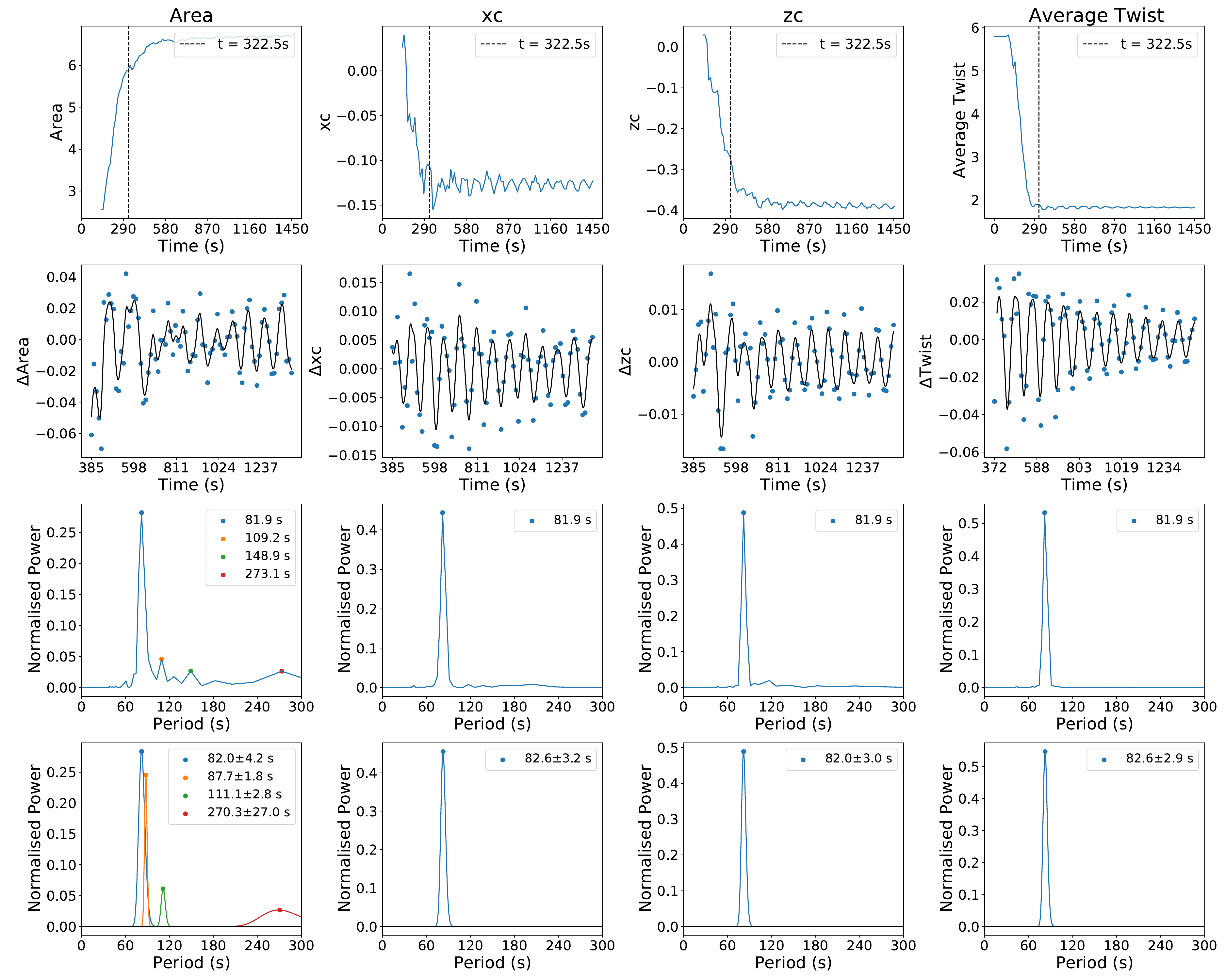}
    \caption[Peak Periods in Structural Oscillations for the Straight Loop]{Analysis of the structural oscillations observed in the straight loop model. The first row illustrates how the fitted ellipses cross-sectional area, central coordinates ($x_c$ and $z_c$), and how the average twist of the loop evolves over time. A dashed black line separates data points before and after 322.5s. In the second row, the post 322.5s data, with the removal of its moving average, is presented. The third row showcases the periods of the power spectrum of this processed data. Finally, the fourth row maps Gaussian peaks to the corresponding peak periods identified in power spectra.}
    \label{fig:straight-structural-oscillations}

    \end{minipage}
  \end{sideways}
\end{figure*}

\begin{figure*}
  \centering
  \begin{sideways}
    \begin{minipage}{\textheight} % Use minipage to limit the width
      \centering
      \includegraphics[width=0.9\textwidth, height = 0.9\textheight, keepaspectratio]{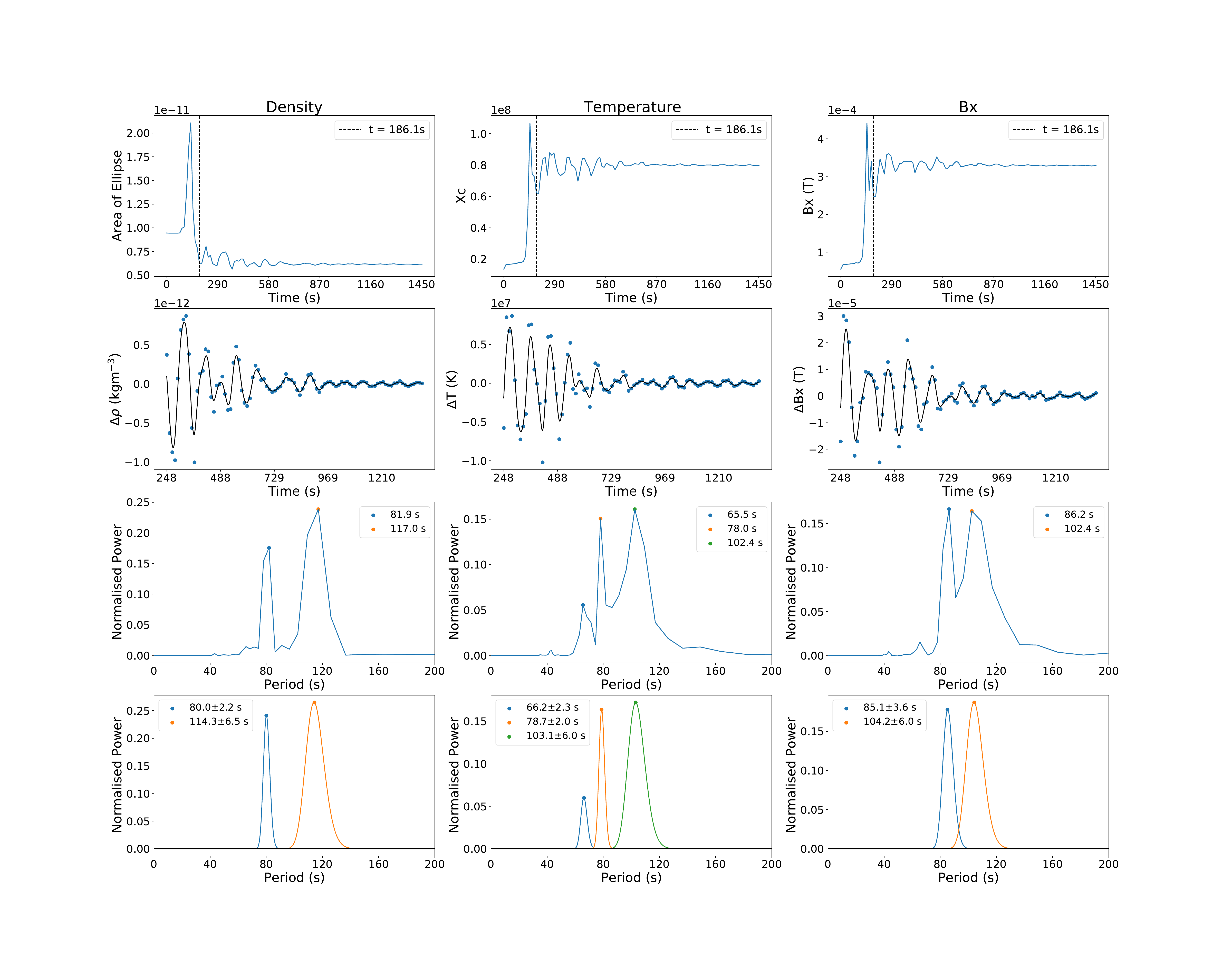}
      \caption[Peak Periods for Plasma Parameter Oscillations for the Straight Loop]{Analysis of parameter oscillations at the centre of the loop. The first column analyses the system's density, the second column examines temperature, and the third column, $B_x$. The rows follow the same structure as those in Figure \ref{fig:straight-structural-oscillations}.}
    \label{fig:straight-density-oscillations}
    \end{minipage}
  \end{sideways}
\end{figure*}

\begin{figure*}
    \centering
    \includegraphics[width = 0.8\textwidth]{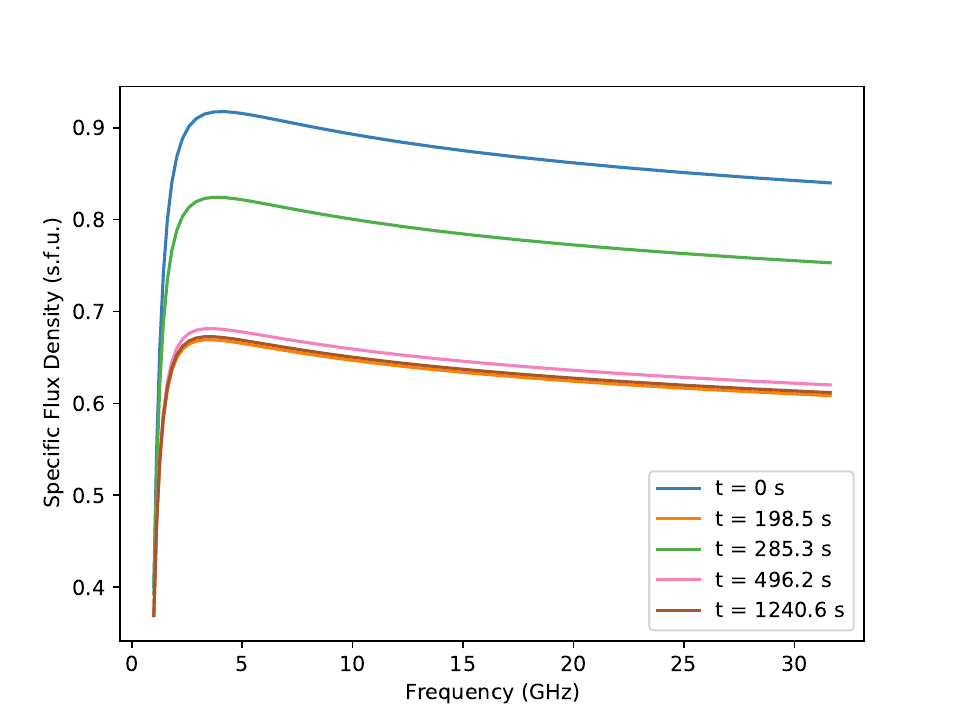}
    \caption[Intensity of GS Radiation vs Frequency for the Straight Loop]{Intensity of GS radiation emitted from the loop-top as a function of frequency at various points in time (s) throughout the straight loop's evolution.}
    \label{fig:straight-gs-radiation-vs-time}
\end{figure*}

\begin{figure*}

      \centering
        \includegraphics[width = \textwidth]{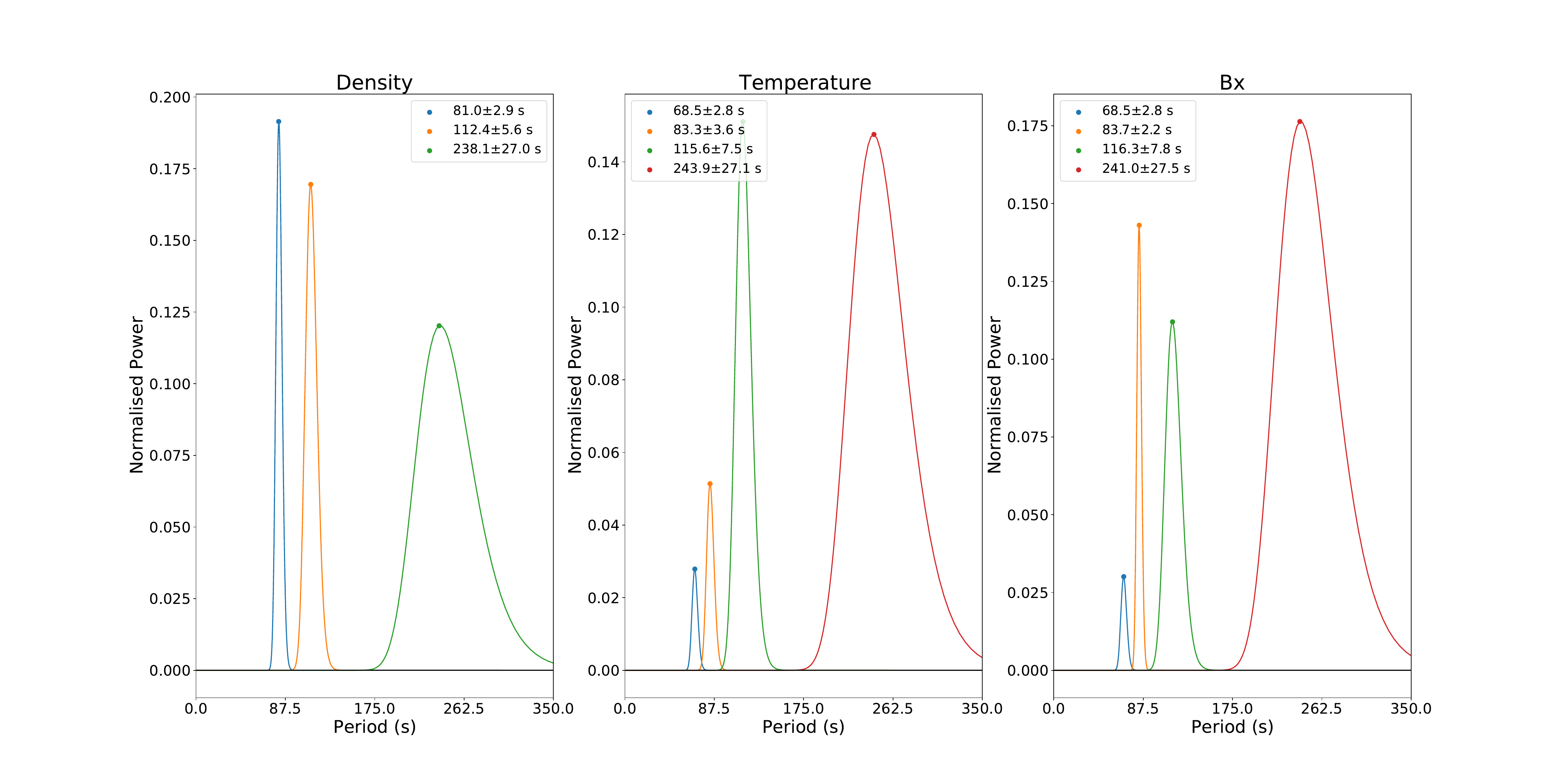}
        \caption{Periods of the power spectrum of for internal parameter oscillations observed at $x = 0.5$, $z = 0.0$ for the straight loop. The first column analyses the system's density, the second column examines temperature, and the third column, $B_x$.}
    \label{fig:alternative_density_oscillations_straight_loop}

\end{figure*}

\begin{figure*}

      \centering
        \includegraphics[width = 1.1\textwidth]{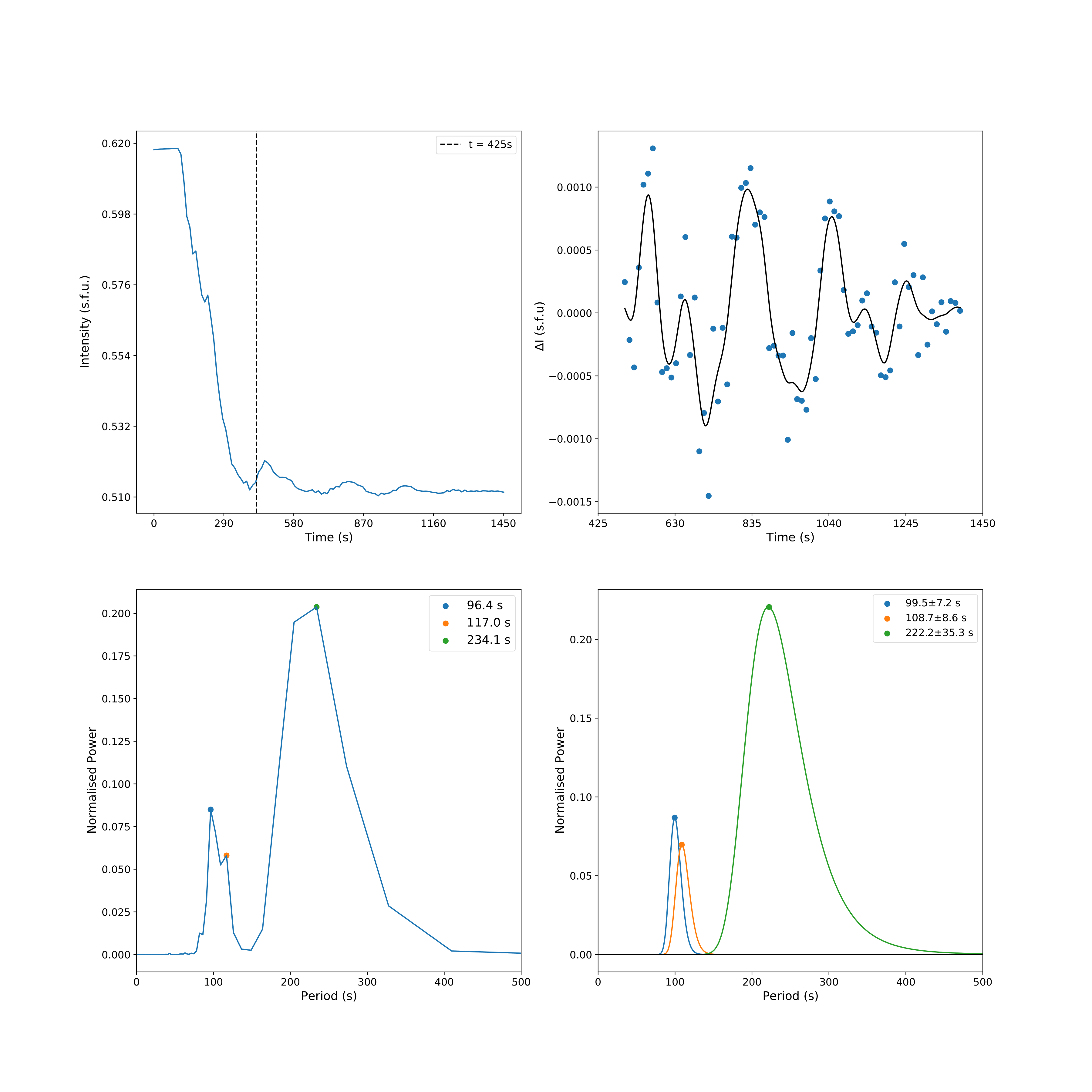}
        \caption[Peak Periods Detected in GS Oscillations for the Straight Loop]{The variation of GS radiation at 1.2 GHz with time (top-left) for the straight loop, detrended GS radiation (top-right), the periods identified in its associated power spectrum (bottom-left), and Gaussian fits to those periods (bottom-right).}
    \label{fig:straight-gs-radiation}

\end{figure*}

\section{Results}\label{Results}

The kink instability was induced and simulated for both straight and curved loops. The resulting evolution, including magnetic reconnection, plasma heating, emitted radiation, and structural oscillations, is described below.

\subsection{Evolution of the Straight Loop}\label{Results:Straight-Loop}

We generate a model of a straight coronal loop in a state of unstable force-free equilibrium, with a twisting parameter $\lambda = 2.3$ as described in Section \ref{LinearModel}. Upon starting the simulation, the loop undergoes the kink instability and begins to reconnect, eventually relaxing to a new equilibrium state.\\

Figure \ref{fig:straight-side-view} shows how the interior magnetic field lines develop as the system evolves. Figure \ref{fig:straight-energetics} depicts the evolution of both the average twist and the energetics of the system over time. We calculated the average twist by determining the total twist of 200 magnetic field lines distributed inside the loop around the central loop axis and then averaging these values. Additional visualisation is provided in Figure \ref{fig:straight-loop-evolution}, which takes a slice of the loop in the X-Z plane at y = 0, and shows how the in-plane magnetic field, density, temperature, and current magnitude within the loop change with time.\\

The simulation was run for 702 Alfvén times (1453s). Between $t = 0$s to around $t = 286$s, inhomogeneities in the in-plane magnetic field, density, temperature, and current magnitude form as the kink instability progresses from its linear phase into its non-linear phase. Reconnection at multiple current sheets within the loop follows and we observe similar magnetic field and energetic evolution to \citet{Hood2009}. After around $t = 286$s, the loop relaxes towards a new equilibrium with reduced twist and magnetic energy; but we do not extend the simulation to reach  a fully static equilibrium.\\

Post-reconnection, we observe multiple structural oscillations in the loop (Figure \ref{fig:straight-structural-oscillations}). We take a cross-sectional slice of the loop (at the midplane $y = 0$) and use our edge-detection and ellipse-fitting algorithm to monitor the changes in the area and central coordinates ($x_c$ and $z_c$) of the loop-top over time (see Section \ref{StructuralOscillation-Calculations}). We observe an expansion of the loop-top, likely driven by Ohmic heating that increases the loop's internal temperature and by reconnection of the loop's field lines with ambient untwisted field lines \citep{Gordovskyy2014}. We also observe lateral shifts in the loop-top in the $x$ and $z$ directions. The central coordinates oscillate around new equilibrium values mainly after about $t = 323$s (though some oscillations  are observed before this point). The expansion and the contraction of the cross-sectional area resemble a sausage-mode oscillation, and the swaying motions detected in $x_c$ and $z_c$, resemble a kink mode oscillation. We also detect clear post-reconnection oscillations in the loop's average twist (see first panel of Figure \ref{fig:straight-energetics}).\\

We analyse these structural oscillations further by removing a moving average from the original data and calculating the oscillations' power spectra. The periods contained within the spectra are then identified. For each oscillation, we observe multiple broad peaks which can be used to determine the dominant periods contributing to each oscillation (see lower panels of Figure \ref{fig:straight-structural-oscillations}). Spectral leakage and inaccuracies are expected when calculating these periods due to limitations in the simulation's total length and temporal resolution. To address this, Gaussian peaks were fitted to each peak. The means of these curves provide a value for peak periods, while the variances serve as errors which allow for comparison with peaks from other power spectra.\\

We detect in each variable a dominant period at $\sim 83$s, along with smaller contributions detected in the sausage-mode oscillations. This common period suggests two things: the structural oscillations are connected, and the oscillations observed in the average twist are likely due to the expansion and contraction of the loop rather than the presence of an additional "torsional-like" Alfv\'en wave, which would be characterised by oscillatory rotational motions around the loop-top.\\

Using a similar methodology we construct power spectra of oscillations detected in the density, temperature, and line-of-sight magnetic field ($B_x$) at the centre of the loop-top [0,0,0] (Figure \ref{fig:straight-density-oscillations}). These variables in a complex and non-linear way together determine the emitted GS radiation. Two common periods are observed in each quantity. The first overlaps with the $\sim 83$s mode identified in the structural oscillations. The second, at $\sim 107$s, aligns with a minor peak observed in the cross-sectional area power spectra. This suggests that the parameter oscillations are primarily associated with the observed structural oscillations. The other minor peaks observed may be associated with harmonics or other modes of oscillations not identified in this paper.\\

Away from the reconnection site, additional oscillations are detected in the density, temperature and line-of-sight magnetic field. Figure \ref{fig:alternative_density_oscillations_straight_loop}) shows that at point $x=0.5$, $z= 0.0$, a new oscillation at approximately $\sim 240s$ emerges alongside the $\sim83$s and $\sim 107$s oscillations. Interference could be a potential reason for why this additional mode does not appear at the centre of the reconnection site.

Oscillations at $\sim 83$s and $\sim 107$s are observed along with a new oscillation at $\sim 240$s at $x=0.5$, $z= 0.0$. All three align with oscillations detected in the sausage-mode.\\

Finally, we identify oscillations in the GS radiation emitted from the loop-top (see Section \ref{GS-Calculations}). To calculate the emitted radiation, we take $\sim$ 9000 lines-of-sight transverse to the loop, in the x-direction, near the midplane, equally spaced within a region bound by [-2:2, -0.67:0.67, -2:2], and record the density, temperature, and line-of-sight magnetic field along those lines. Subsequently, we partition this region into 356 sub-regions, and for each sub-region, we average the aforementioned parameters for each point between each line-of-sight to reduce the computational expense of the GS calculations. We avoid taking an average of the whole loop-top as different segments of the loop contribute different quantities of GS radiation due to varying internal parameters throughout the simulation. We carefully select a number of sub-regions that strike a balance between reducing computational strain and ensuring that the final radiation calculated remains representative of the behaviour of the loop.\\

The GS microwave frequency spectra emitted from this region at different points in time are illustrated in Figure \ref{fig:straight-gs-radiation-vs-time}. Continuous emission is predominantly observed around 1 GHz, with the total intensity 
decreasing over time as the magnetic field strength of the loop weakens. The spectra show typical shapes, with optically-thick radiation at low frequencies rising to a peak, with gradual decrease  of intensity through  the higher (optically-thin) frequency range. We also detect oscillations across the calculated spectrum, similar to \citet{Smith2022}. In Figure \ref{fig:straight-gs-radiation}, we focus on the radiation emitted at 1.2 GHz noting that other frequencies exhibit similar behaviour.
This choice of frequency for analysis means we are in the partially optically-thick regime. This means the microwave emission may have a less direct correspondence with the underlying plasma and magnetic field parameters than in the optically-thin regime .\\

%It is likely this is a harmonic of the $107$ s mode. This is consistent with simulations of GS emission from an idealised slab model of kink oscillations \citep{Kaltman_2023}, which find that the oscillation frequency of GS emissions from kink modes at the antinode (which would correspond to the loop-top) is twice that of the underlying kink. 

A primary peak is observed at $\sim 222$s, with two secondary peaks around $107$s. These peaks align with those found in the internal parameter oscillations and also appear as minor peaks in the sausage-mode oscillations. Notably, no peak is observed at $\sim83$s.\\

There is a complex relationship between the sausage-mode oscillations, internal parameter oscillations, and oscillations in the emitted GS radiation. This complexity is expected due to the highly non-linear mechanisms by which GS emission arises from plasma parameters and the magnetic field \citep{Mossessian2012, Kupriyanova_2022, Kaltman_2023}. A notable feature is that though the $~222.2$ s mode is dominant in the emitted GS radiation and internal parameter oscillations (away from the centre loop-top), it is less prominent in structural oscillations. There are several potential explanations for this. \\

One possibility is that the $222.2 \pm 35.2$ s GS oscillation results from interference between the $111.28 \pm 2.8$ s and $82.0 \pm 4.2$ s peaks seen in the sausage-mode oscillations. Alternatively, other wave modes—undetected by the edge-detection algorithm could contribute to the emission oscillations. A potential candidate is a longitudinal acoustic mode, whereas a torsional mode can be ruled out since the $\sim222$ s peak would also appear in the average twist oscillations, which was not observed. Furthermore, since the radiation we analyse is in the optically-thick region of the spectrum, it will be determined by the plasma and magnetic field across the line-of-sight in a complex way, and not just depend on the local conditions at the emission site. 

\begin{figure*}
  \centering
  \begin{subfigure}[b]{0.47\textwidth}
    \centering
    \includegraphics[width=\linewidth]{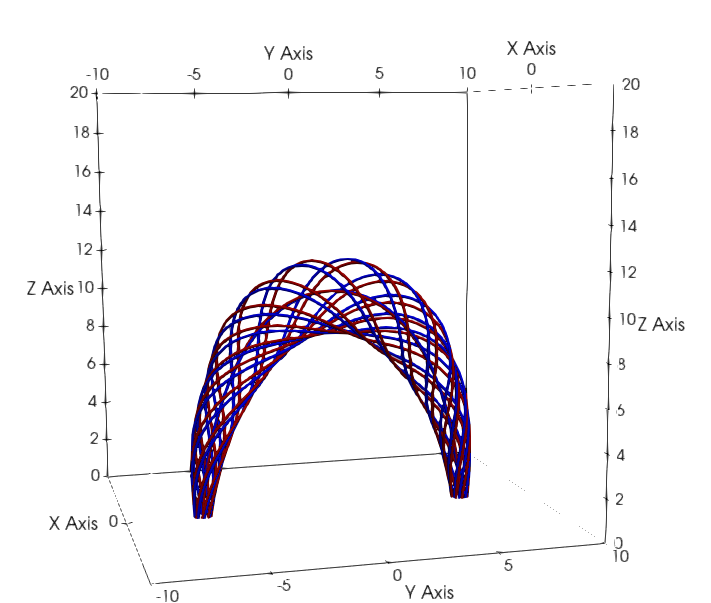}
    \caption{$t = 545$ s}
    \label{fig:subfig-a}
  \end{subfigure}
  \hfill
  \begin{subfigure}[b]{0.47\textwidth}
    \centering
    \includegraphics[width=\linewidth]{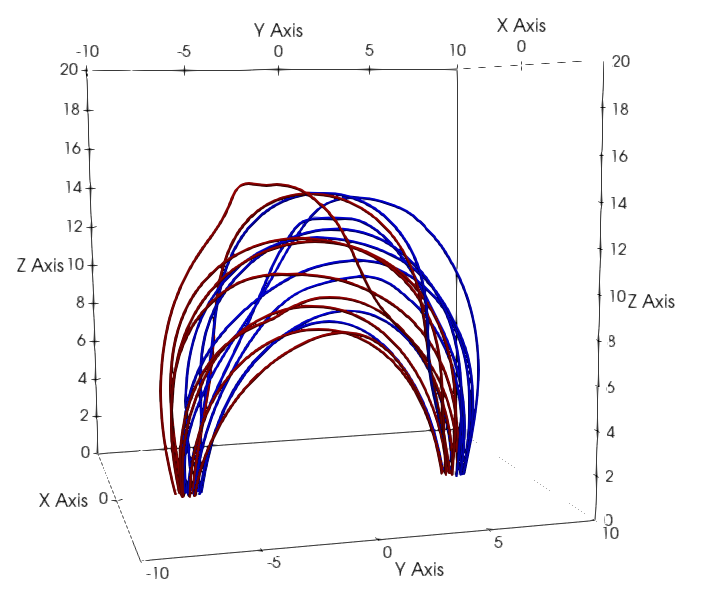}
    \caption{$t = 725$ s}
    \label{fig:subfig-b}
  \end{subfigure}
  
  \vspace{1cm} % Adjust vertical spacing between rows
  
  \begin{subfigure}[b]{\textwidth}
    \centering
    \includegraphics[width=0.47\linewidth]{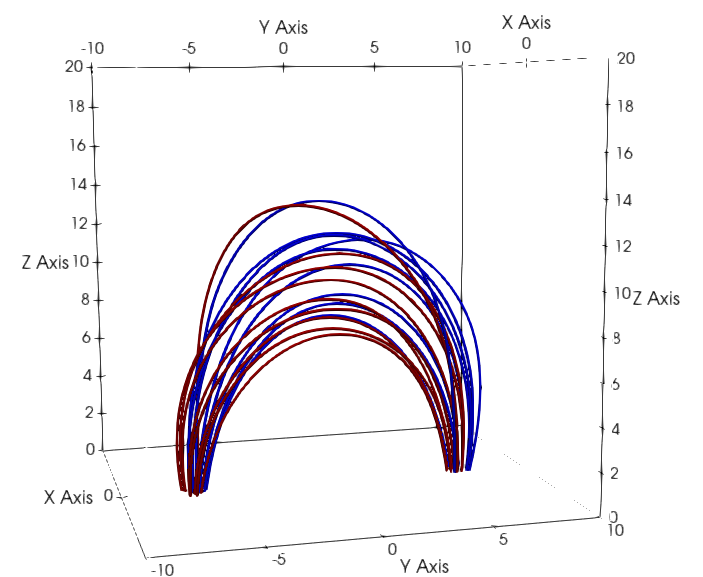}
    \caption{$t = 1035$ s}
    \label{fig:subfig-c}
  \end{subfigure}
  
  \caption{Evolution of the curved model coronal loop's magnetic field lines in the Y-Z plane over time. The blue field lines originate from the foot-point at $y = -6.4$, while the red field lines originate from the foot-point at $y = 6.4$.}
  \label{fig:curved-side-view}
\end{figure*}

\begin{figure*}
    \centering
    \includegraphics[width = \textwidth]{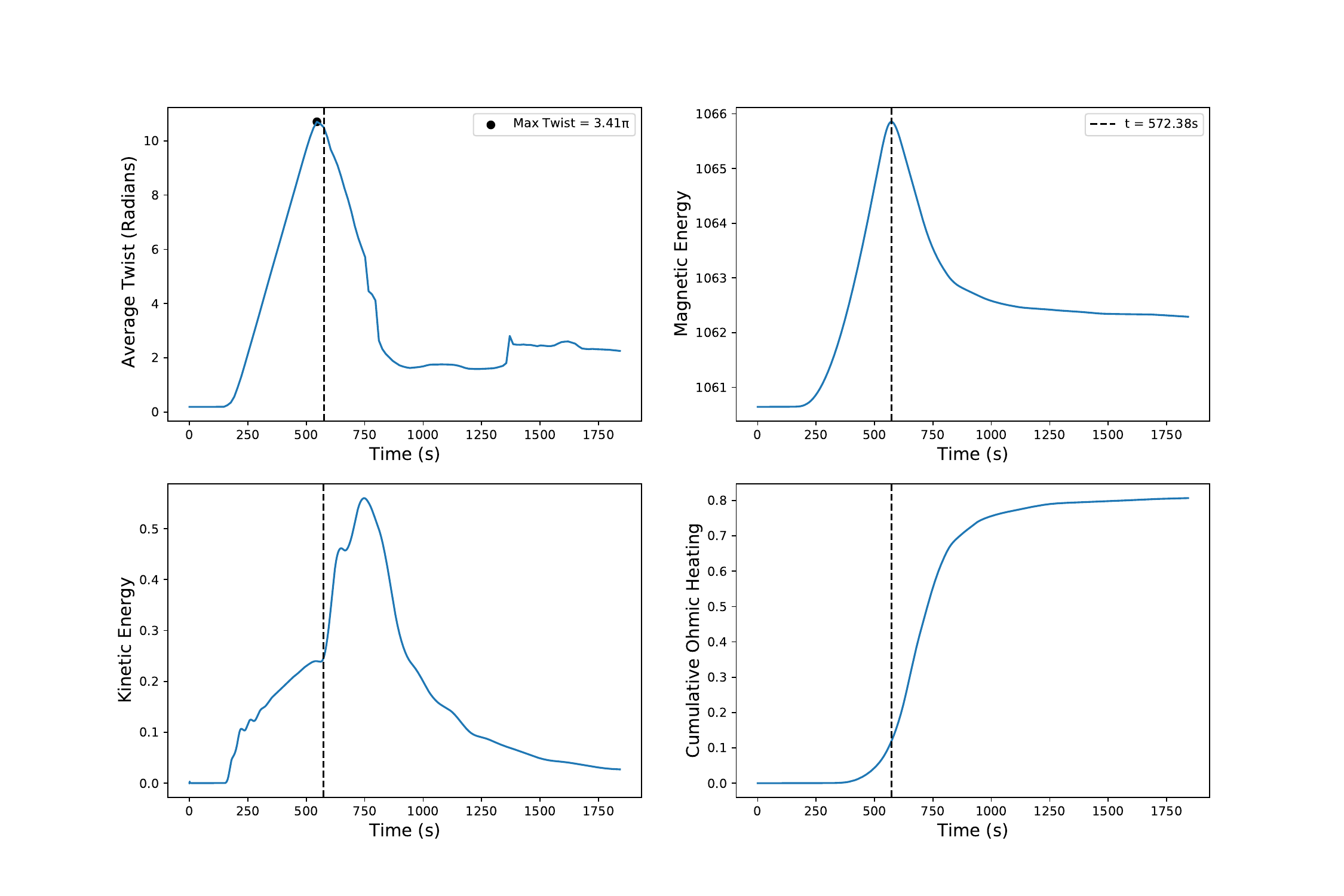}
    \caption[Evolution of Energetics and Average Twist in the Curved Loop vs. Time]{The temporal evolution of: average twist of the loop (top-left), total magnetic energy (top-right) and kinetic energy (bottom-left) of the system, and cumulative Ohmic heating (bottom-right). A black dashed line has been separates the initial setting up of the initial state and the resultant evolution.}
    \label{fig:curved-energetics}
\end{figure*}

%\begin{figure*}
    %\centering
    %\includegraphics[width = \textwidth, height = \textheight,keepaspectratio]{Figures/Kink_Figures/loop-evolution.pdf}
    %\caption[Evolution of Temperature at the Curved Loop Midplane vs. Time]{Evolution of temperature of the curved loop midplane over time.}
    %\label{fig:curved-loop-evolution}
%\end{figure*}

\begin{figure*}
  \centering
  \begin{sideways}
    \begin{minipage}{\textheight}

        \centering
    \includegraphics[width = 0.8\textwidth]{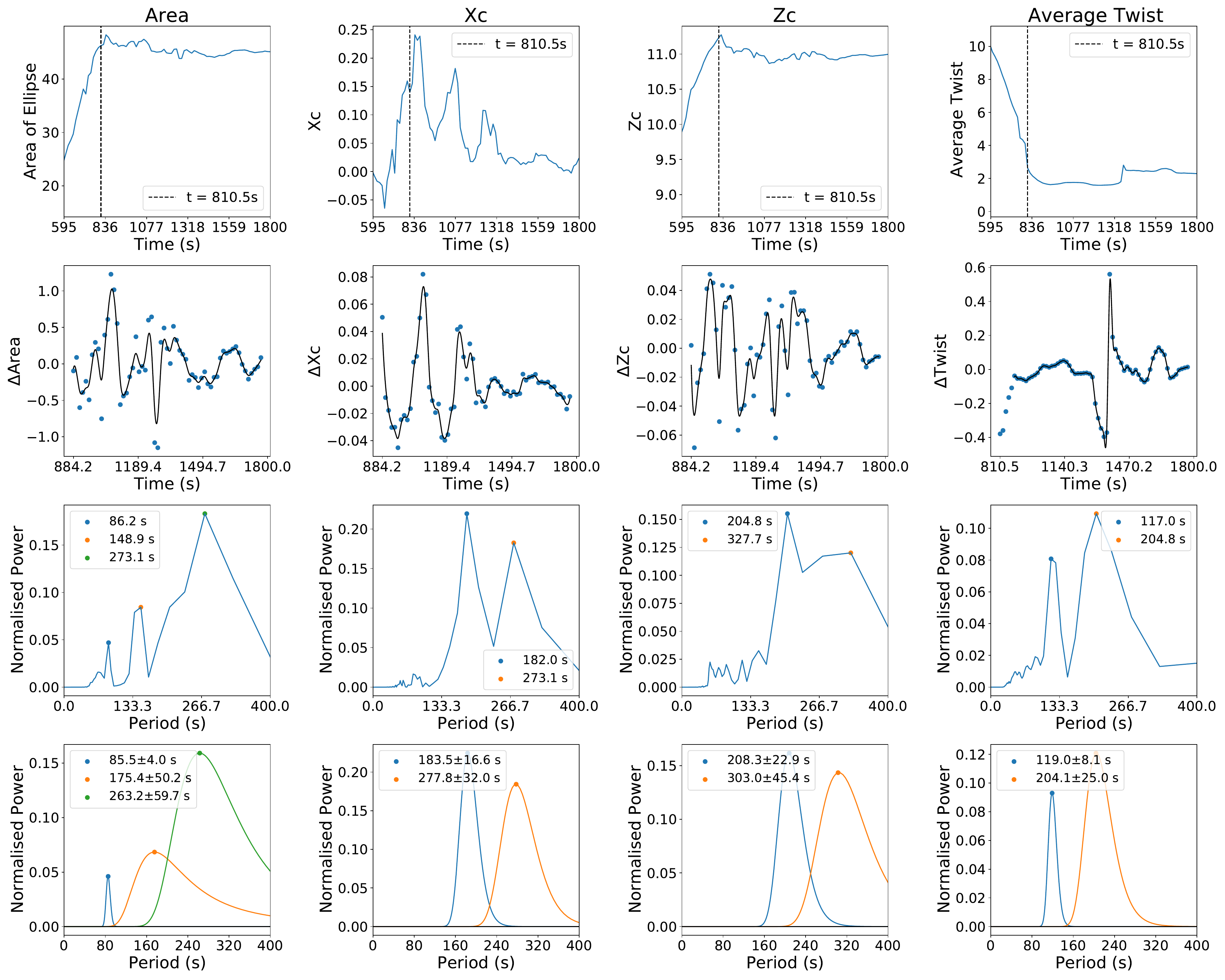}
    \caption[Peak Periods in Structural Oscillations for the Curved Loop]{Analysis of the structural oscillations observed in the curved loop model, following the same structure as Figure \ref{fig:straight-structural-oscillations}.}
    \label{fig:curved-structural-oscillations}
    \end{minipage}
  \end{sideways}
\end{figure*}

\begin{figure*}
  \centering
  \begin{sideways}
    \begin{minipage}{\textheight}
        \centering
    \includegraphics[width = 0.95\textwidth]{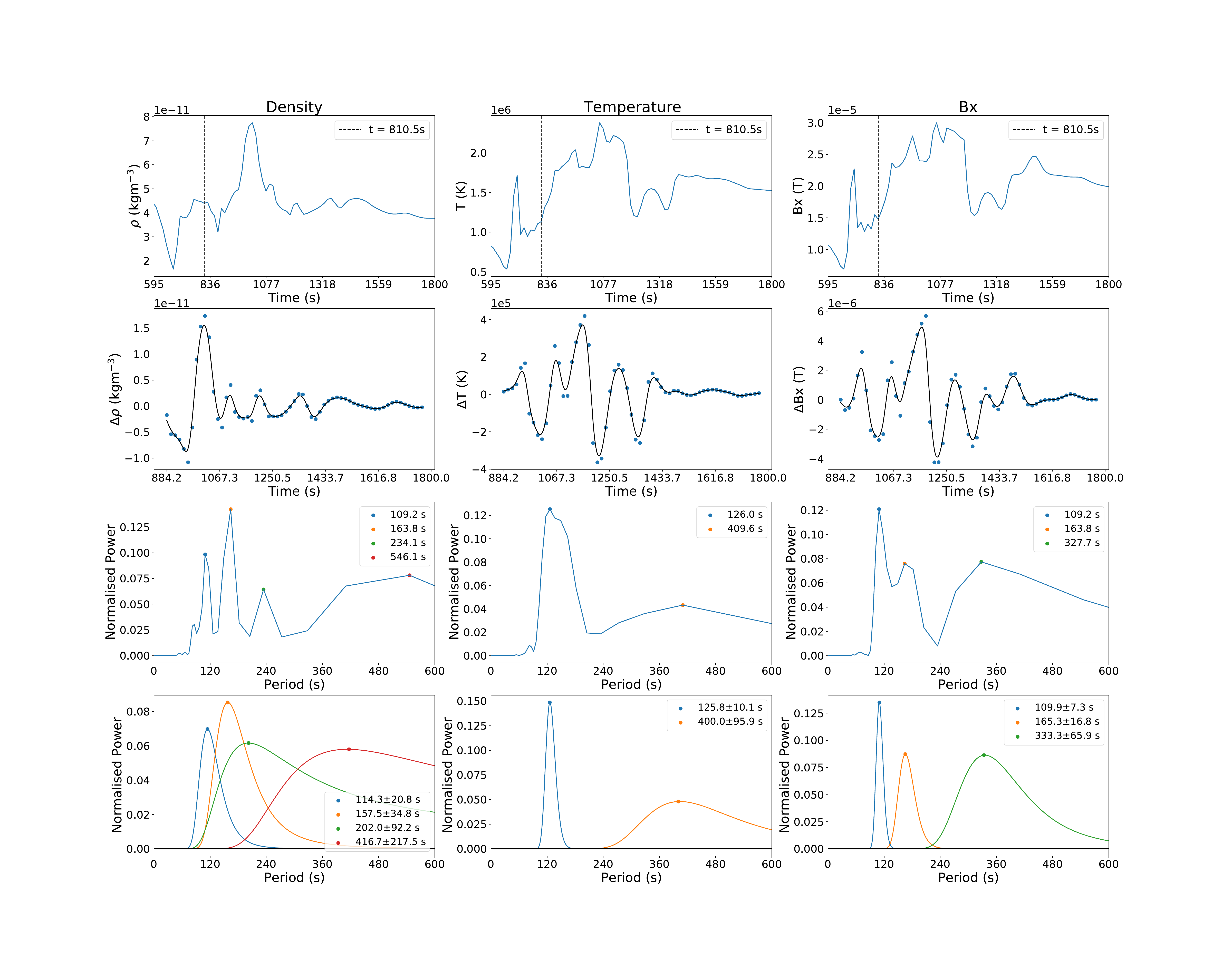}
    \caption[Peak Periods for Plasma Parameter Oscillations for the Straight Loop]{Analysis of parameter oscillations at [0,0,12], located at the midplane of the curved loop. The first column analyses the system's density, the second column examines temperature, and the third column, $B_x$. The rows follow the same structure as those in Figure \ref{fig:straight-density-oscillations}.}
    \label{fig:curved-density-oscillations}
    \end{minipage}
  \end{sideways}
\end{figure*}

%\begin{figure*}

      %\centering
        %\includegraphics[width = \textwidth]{alternative_density_oscillations_analysis.pdf}
        %\caption{\textbf{INSERT TEXT HERE}}
    %\label{fig:alternative_density_oscillations_analysis_curved_loop}

%\end{figure*}

%\begin{figure*}
    %\centering
    %\includegraphics[width = \textwidth]{Figures/Kink_Figures/curved_gs_radiation_vs_time.pdf}
    %\caption[Intensity of GS Radiation vs Frequency for the Curved Loop]{Intensity of GS radiation emitted from the loop-top as a function of frequency at various points in time throughout the curved loop's evolution}
    %\label{fig:curved-gs-radiation-vs-time}
%\end{figure*}

\begin{figure*}
        \centering
    \includegraphics[width = 0.9\textwidth]{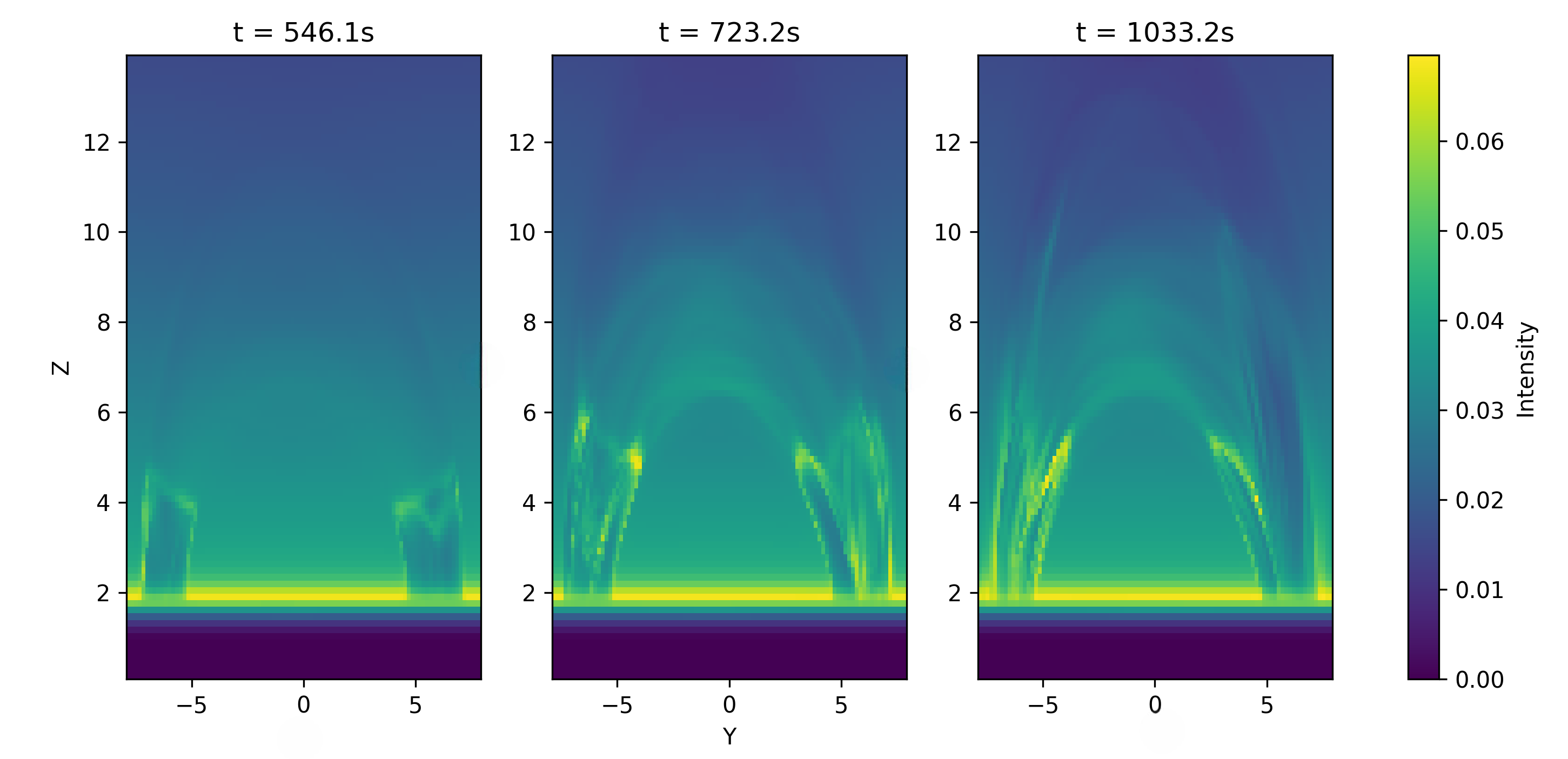}
    \caption{
   Normalised gyrosynchrotron radiation at 15 GHz in the Y–Z plane (spanning Y = –8 to 8, Z = 0 to 14), for the same timesteps as images of the evolving curved coronal loop presented in Figure} \ref{fig:curved-side-view}.
    \label{fig:curved-gs-map}

\end{figure*}

\begin{figure*}
        \centering
    \includegraphics[width = \textwidth]{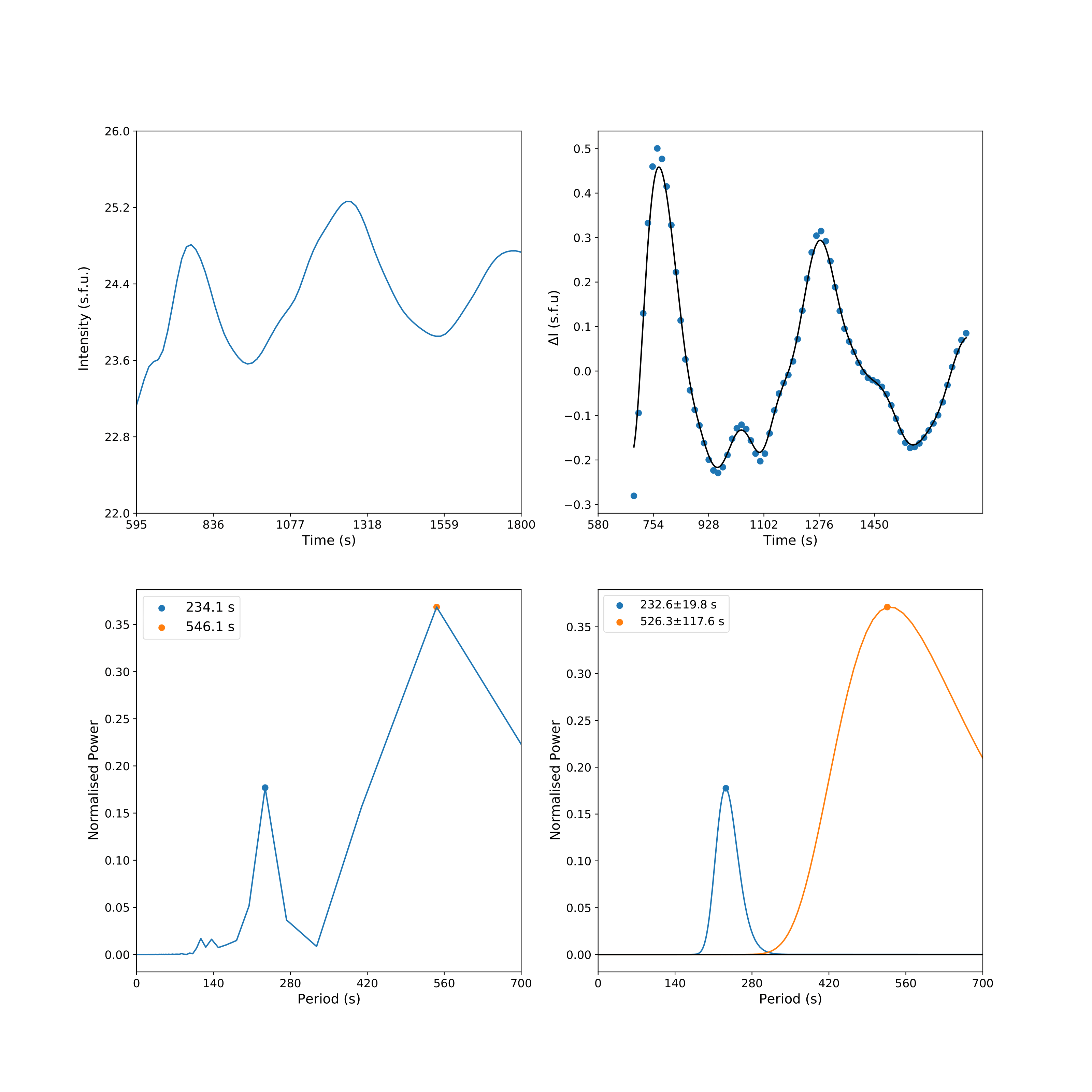}
    \caption[Peak Periods Detected in GS Oscillations for the Curved Loop]{
    Analysis of the measured gyrosynchrotron radiation at 15 GHz for the curved loop. The first row demonstrates how the GS radiation for each frequency evolves over time. The subsequent rows follow the same structure as Figure \ref{fig:straight-gs-radiation}.}
    \label{fig:curved-gs-radiation}

\end{figure*}

\subsection{Evolution of the Curved Loop}\label{Results:Curved Loop}

Following \citet{Gordovskyy2014}, we induce the kink instability in a curved loop by applying a twist to the loop's foot points. We inject a total twist of approximately $4\pi$ before the loop becomes unstable, reconnects, and relaxes towards a new equilibrium state.\\ 

%%When considering the results of these simulations it is important to remember that the model is only physically representative of what happens in the solar corona after we turn off the twisting profile ($\sim t = 565$s).\\

We analyse the curved loop in a similar way to the straight loop. The simulation was run for 1500 Alfvén times (1845s). Once the loop reaches an average twist of $3.41\pi$, at around $t = 565$s, the loop becomes unstable, indicating that some of the injected twist dissipated during this initial phase. The evolution prior to this is only to establish an unstable equilibrium, and this time corresponds to $t=0$ for the straight loop discussed above. When the loop begins to reconnect, it exhibits similar behaviour to the straight loop concerning energetics and internal dynamics (see Figures \ref{fig:straight-energetics}-\ref{fig:straight-loop-evolution}) and is similar to previous work with this model \citep{Gordovskyy2014,Bareford_2015,Pinto2016,Smith2022}.\\

The evolution of the loop's interior magnetic field lines is depicted in Figure \ref{fig:curved-side-view}, while Figure \ref{fig:curved-energetics} illustrates the evolution of the average twist and energetics over time. The internal dynamics of the loop evolve similarly to the straight loop (Figure \ref{fig:straight-loop-evolution}). Discussions of behaviour prior to approximately $t = 565$ seconds will be omitted, as this phase serves solely to set up an unstable twisted approximate equilibrium and the dynamics are non-physical.\\

We again use our edge-detection and ellipse-fitting algorithm on a cross-sectional slice of the loop at $y = 0$ to quantify changes in the loop-top's area and central coordinates ($x_c$ and $z_c$) over time (see Figure \ref{fig:curved-structural-oscillations}). We observe expansion and contraction of the loop, similar to the straight loop, and lateral displacement. The direction of the lateral shift differs in the curved loop. This is due to the straight loop's symmetry, which results in a random initial shift, whereas the curvature of the curved loop provides a preferred direction. In Figure \ref{fig:curved-structural-oscillations}, we also identify post-reconnection oscillations in the loop's average twist. \\

A single common period of approximately 190 seconds is observed in all four structural parameters. Additionally, a second period of about 270 seconds is observed in all parameters except the average twist. This suggests, similar to the straight loop, that the sausage-mode and kink-mode oscillations are connected, and the sausage-mode drives oscillations in the loop's twist. However, unlike the straight loop, a unique period is observed in the average twist oscillations ($\sim 119$s), indicating the presence of a potential alternative mechanism that is also affecting the loop's twist.\\

Further oscillations in the density, temperature, and line-of-sight magnetic field are observed inside the loop-top over time, measured at coordinates [0,0,12] (Figure \ref{fig:straight-density-oscillations}). This point is located internally at the centre of the loop, drifts from the centre as the loop evolves but remains within the loop-top throughout the simulation.  
A peak at $\sim 190$s is observed in the density and $B_x$ oscillations, while a peak at $\sim 270$s is observed in all three variables. These two peaks are associated with the sausage-mode and kink-mode oscillations. 
A third peak at $\sim 119$s is also observed in all three variables. This peak aligns with the 119-second peak observed in the loop's average twist, suggesting that the internal parameters are also affected by the mechanism influencing the loop's average twist, perhaps torsional Alfv\'en waves.\\

To calculate the GS radiation, we use the same approach as the one used for the straight loop. However, this time we consider 18,350 lines of sight along the x-axis, bounded by [-10:10, -2:2, 7:14], divided into 356 sub-regions. Compared to the straight loop, we observe a wider range of emitted frequencies, resembling those seen in \citet{Smith2022}.\\

%Figure \ref{fig:curved-gs-radiation-vs-time} illustrates the GS radiation emitted from this region as a function of frequency at different points in time. 

%Compared to the straight loop, we observe a wider range of emitted frequencies, resembling those seen in \citet{Smith2022}. The intensity of the loop's emissions increases as the loop becomes twisted and decreases similarly tothe straight loop upon the onset of the kink instability.\\

The intensity vs. frequency distribution for the curved loop has the same shape as the straight loop (Figure \ref{fig:straight-gs-radiation-vs-time}), with a peak at $15$ GHz. Figure \ref{fig:curved-gs-map} shows a map of gyrosynchrotron radiation emitted from the coronal loop in the Y–Z plane at 15 GHz, at the same snapshots as in Figure \ref{fig:curved-side-view}. A total of 146,800 lines of sight are grouped into 9,025 averaged sub-regions, from which the emitted radiation is calculated. Brightenings and fine structure are observed within the loop. While this level of spatial resolution cannot be observed with current observational techniques, the figure provides insight into the origins of the emission peaks.\\

Figure \ref{fig:curved-gs-radiation} focuses on oscillations at $15$ GHz, noting that other frequencies exhibit similar behaviour. We observe two peaks: one at approximately $233$ seconds and another around $526$ seconds. The first peak aligns with the $270$-second oscillations observed in the structural modes, suggesting that, similar to the straight loop, it is driven by sausage and kink mode oscillations.\\

However, the dominant peak at 526 seconds does not directly correspond to any structural oscillations. As in the straight loop, there is evidence to suggest that they may arise from interference effects. Interference of the peaks at $175.4 \pm 50.2$ and $263.2 \pm 59.7$s area oscillations, $183.5 \pm 16.6$ and $277.8 \pm 32.0$s in the $x_c$ oscillations, and $208.3 \pm 22.9$ and $303.3 \pm 45.4$s in the $z_c$ oscillations give a period similar to the 526-second peak.  Additionally, interference between adjacent peaks in the density oscillations also aligns with the 526-second peak. We also note that the 119-second oscillation detected in the loop's average twist does not appear in the GS radiation oscillations. Though this mechanism remains unidentified, it does not significantly influence the oscillations in the emitted GS radiation.

\section{Discussion and Conclusions}\label{Discussion}

We performed 3D resistive MHD simulations of both  a straight and curved kink-unstable twisted coronal loop using LARE3D. Both loops underwent reconnection, releasing stored magnetic energy,  and exhibited broadly similar behaviour in terms of their energetics and internal dynamics. The GS emissions, observed in the microwave band, were forward-modelled using a fast GS code developed by \citet{Fleishman2010}, similar to the approach used in \citet{Smith2022}. The outcomes  of our study are two-fold. Firstly, we have demonstrated a new methodology for analysing structural oscillations in a realistic model of a solar coronal loop, and used this to identify these. Secondly, we have explored the relationship between these structural modes, and the associated local oscillations in plasma and magnetic field parameters, with oscillations in the GS emission light-curves.  \\

Using new methodology, we identified "structural" oscillations of the loop (kink, and sausage mode oscillations) by taking a cross-sectional slice of the loop-top, fitting an ellipse using Canny edge detection, Delaunay triangulation and alpha shapes, and tracking the evolution of the loop-top's shape. We calculated the peak periods in the power spectra of these structural oscillations and compared them to peak periods observed in the emitted GS radiation, average twist, and internal parameters of the loop. \\

We identified sausage-mode and kink-mode oscillations in both loops. These oscillations shared similar peak periods, which were also observed in the oscillations of the loop's average twist. A potential reason for this is that a swaying loop, moving through regions of increased and decreased magnetic field strength, would lead to the loop-top periodically expanding and contracting. This would manifest as a sausage-mode oscillation and would also induce oscillations in the loop's average twist.\\

For the straight loop, a dominant period at approximately 222 seconds was observed in the oscillations of the GS emissions. Two similar peaks were also observed around 107 seconds. The 107-second period aligns with minor peaks found in the sausage-mode power spectra and oscillations seen in the internal parameters of the loop-top. The 222-second peak was not directly associated with any structural oscillation, which is expected due to the strongly non-linear relationship between radiative transport and radiative emissions. It was observed in the internal parameter oscillations and interference between observed sausage-mode oscillations gives a period that aligns with the 222-second peak. Alternatively, other wave modes—undetected by the edge-detection algorithm could contribute to the emission oscillations. The approximately 83-second period observed in the structural and internal parameter oscillations did not appear in the GS emissions at all.\\

Similar conclusions can be drawn for the curved loop. The peaks in the GS emissions were observed at $\sim 233$s and $\sim 527$s. The first peak overlapped with the 270-second peaks observed in the structural oscillations, indicating that like the straight loop, one of the peaks in the GS oscillation is also driven by the sausage and kink mode oscillations. The dominant $\sim 526$s peak is not associated with any structural oscillations but could be associated with interference in the sausage-mode, kink-mode and density oscillations. This suggests that for both the straight and curved loop, that the GS oscillations could be generated by sausage and kink-mode oscillations and interference of those modes.\\

A potential additional process  was observed in the curved loop, affecting the oscillations in the loop's average twist. This oscillation had a period of approximately $119$s and could be caused by a torsional-like mode, which would induce oscillations in the average twist of the loop due to rotational motion around the loop-top. However, the 119-second oscillation did not appear in the GS radiation oscillations, suggesting that even if an additional mechanism was affecting the loop's twist, it did not contribute to the dominant period in the GS oscillations. Furthermore, in the straight loop, only one period was observed in the average twist oscillations and this was shared with the periods observed in the kink and sausage mode oscillations. This suggests that there were no other mechanisms driving the oscillations in the average twist for the straight loop, discounting the potential contribution of a torsional-like mode for the straight loop as well.\\

In summary, both sausage and kink-mode oscillations were detected in the straight and curved loop models. While these modes play a crucial role in shaping the GS emission—and thus any observed QPPs in the flare—the relationship between light-curve pulsations and underlying loop oscillations remains complex. This is not unexpected, due to the non-linear dependence of GS emission on the plasma and field parameters \citep{Mossessian2012,Kupriyanova_2022,Kaltman_2023}. Optically-thick effects on the radiation may play a significant role in the relationship between the parameters in the emission regions and in the observed radiation. It is possible that the sausage and kink-mode oscillations are the sole contribution to the GS oscillations, but it is also possible that mechanisms unidentified here play a role in determining the frequencies of the GS oscillations.\\

There is evidence to suggest that sausage and kink-mode oscillations (and interference of these modes) may account for the GS oscillations emitted from the straight and curved loop model. While a torsional mode is unlikely to be responsible, other unidentified mechanisms could also influence their frequencies. In the curved case, curvature effects could introduce an acoustic mode.\\

Future research should investigate whether GS oscillations result from interference between sausage and kink modes or from another mechanism entirely. If the GS oscillations result from interference, it is important to gain a deeper understanding of how the sausage and kink modes interact, how this interaction translates into observable features of the MW oscillations, and how variations in plasma parameters influence this behaviour. Deepening our understanding of these mechanisms not only increases our understanding of the time-dependent nature of solar flares but paves the way for the development of seismological tools that could be used to determine plasma parameters within a flaring region from observed QPP data in the future.

\section*{Acknowledgements}

We wish to thank the Science and Technology Facilities Council (STFC) for providing studentship support for JS. PKB and MG were
funded by the STFC grant ST/T00035X/1. We also thank the Distributed Research utilising Advanced Computing (DiRAC) group for providing the computational facilities used to run the simulations in this paper. 

%%%%%%%%%%%%%%%%%%%%%%%%%%%%%%%%%%%%%%%%%%%%%%%%%%
\section*{Data Availability}

The data underlying this article will be shared on reasonable request to the corresponding author (JS).



\bibliographystyle{mnras}
\bibliography{bibliography} % if your bibtex file is called example.bib

%\begin{thebibliography}{99}
%\end{thebibliography}

%%%%%%%%%%%%%%%%%%%%%%%%%%%%%%%%%%%%%%%%%%%%%%%%%%

%%%%%%%%%%%%%%%%% APPENDICES %%%%%%%%%%%%%%%%%%%%%

%\appendix



%%%%%%%%%%%%%%%%%%%%%%%%%%%%%%%%%%%%%%%%%%%%%%%%%%

% Don't change these lines
\bsp	% typesetting comment
\label{lastpage}
\end{document}